\documentclass[12pt]{article}

\usepackage{graphicx}
\usepackage{amssymb}
\usepackage{bm}

\usepackage{color}
\usepackage{lineno}

\usepackage{url}

\usepackage{setspace}
\usepackage[margin=1in]{geometry}

\newcommand{\ozone}{O$_3$ }
\newcommand{\ozonens}{O$_3$}

\newcommand{\s}{\mathbf{s}}

\usepackage{subcaption}

\usepackage[colon, square]{natbib}
\title{Distributions of Human Exposure to Ozone During Commuting Hours in Connecticut using the Cellular Device Network}

 \author{Owais Gilani\thanks{Corr. Author: Tel: +1 570 577 1391; Email: owais.gilani@bucknell.edu} \\ {\small Bucknell University, Lewisburg PA, USA} \and  Simon Urbanek \\ {\small AT\&T Labs Research, Bedminster NJ, USA} \and Michael J. Kane \\  {\small Yale University, New Haven CT, USA}}

\begin{document}

\maketitle

%% use the tnoteref command within \title for footnotes;
%% use the tnotetext command for the associated footnote;
%% use the fnref command within \author or \address for footnotes;
%% use the fntext command for the associated footnote;
%% use the corref command within \author for corresponding author footnotes;
%% use the cortext command for the associated footnote;
%% use the ead command for the email address,
%% and the form \ead[url] for the home page:
%%
%% \title{Title\tnoteref{label1}}
%% \tnotetext[label1]{}
%% \author{Name\corref{cor1}\fnref{label2}}
%% \ead{email address}
%% \ead[url]{home page}
%% \fntext[label2]{}
%% \cortext[cor1]{}
%% \address{Address\fnref{label3}}
%% \fntext[label3]{}

%% use optional labels to link authors explicitly to addresses:
%% \author[label1]{Owais Gilani}
%% \address[label1]{Bucknell University, Lewisburg PA, USA}
%% \author[label2]{Simon Urbanek}
%% \address[label2]{AT\&T Labs Research, Bedminster NJ, USA}
%% \author[label3]{Michael J. Kane}
%% \address[label3]{Yale University, New Haven CT, USA}

%%\author{Owais Gilani}

%%\address{Lewisburg, United States}

%% Start line numbering here if you want

\noindent{\bf Abstract:} Epidemiologic studies have established associations between various air pollutants and  adverse health outcomes for adults and children. Due to high costs of monitoring air pollutant concentrations for subjects enrolled in a study, statisticians predict exposure concentrations from spatial models that are developed using concentrations monitored at a few sites. In the absence of detailed information on when and where subjects move during the study window, researchers typically assume that the subjects spend their entire day at home, school or work. This assumption can potentially lead to large exposure assignment bias. In this study, we aim to determine the distribution of the exposure assignment bias for an air pollutant (ozone) when subjects are assumed to be static as compared to accounting for individual mobility. To achieve this goal, we use cell-phone mobility data on approximately 400,000  users in the state of Connecticut during a week in July, 2016, in conjunction with an ozone pollution model, and compare individual ozone exposure assuming static versus mobile scenarios. Our results show that exposure models not taking mobility into account often provide poor estimates of individuals commuting into and out of urban areas: the average 8-hour maximum difference between these estimates can exceed 80 parts per billion (ppb). However, for most of the population, the difference in exposure assignment between the two models is small, thereby validating many current epidemiologic studies focusing on exposure to ozone. \\

%Research question: Should we take into account human mobility when determining ozone exposure for epidemiological studies?

%% main text
\section{Background and Motivation}
\label{S:1}

Epidemiologic studies have established associations between various air pollutants and a number of adverse health outcomes for adults and children. Air pollutants, such as ground-level ozone (\ozonens), particulate matter, carbon monoxide, lead, sulfur dioxide, and nitrogen dioxide, have been shown to worsen health outcomes such as heart rate variability, cardio-pulmonary mortality, acute myocardial infarction, low birth weight, development and exacerbation of asthma, reduced lung function, and acute respiratory symptoms \citep{pope2002, pope2004, peel2005, zanobetti2005, chen2006, brauer2007, delamater2012, pedersen2013, sacks2014}. 

These associations between exposure to air pollutants and adverse health outcomes have been established using various epidemiologic study designs. Some studies have been conducted on an ecological scale, where the unit of analysis is at an aggregated level (such as counties or census tracts). In these studies, an aggregate measure of a health outcome (such as cause-specific mortality rate over time) is correlated with an aggregate measure of pollutant exposure over the study area \citep{peel2005, chen2006, delamater2012,  sacks2014}. While these studies are useful in determining associations on the population level, they are subject to the ecological fallacy so that conducting inference on an individual level is problematic. Other studies have utilized case-control or cohort designs (prospective and retrospective), that have conducted the analysis on an individual scale \citep{gent2003, brauer2007}. In such studies, health data are available on individuals (often in great detail), which are then correlated with data on individual level pollutant exposure. However, assigning pollutant concentration exposure to individuals is rather challenging. Giving each study participant an air pollution monitor to carry with them all day during the study duration is extremely expensive and impractical. Therefore, epidemiologic studies analyzing the effects of air pollutants on adverse health outcomes at an individual level typically estimate pollutant concentrations at subject residences or work places using various approaches (that vary in their level of sophistication)\citep{pope2002, gent2003, jerrett2005, brauer2007, pedersen2013}. 

A common approach to estimate pollutant concentration at an area of interest (e.g. subject residence) uses observed or monitored air pollutant concentration data at limited sites and time durations to predict concentrations at unsampled sites and time durations. This prediction may be achieved using some very simple approaches such as assigning the same concentration as the closest monitored site or an average from a few closest sites \citep{rage2009}, or estimating via techniques such as inverse distance weighting \citep{neidell2004}. More sophisticated approaches include land use regression models \citep{turner2016} and spatial/temporal interpolation techniques using geostatistical models of varying complexity, e.g. universal kriging \citep{jerrett2005, rage2009}. 

Another popular approach for assigning pollutant concentrations to subject residences is using some deterministic computer model output that simulates either the underlying pollutant chemistry (e.g. the Community Multiscale Air Quality or CMAQ model \citep{byun2006}), or the dispersion process of air pollutants from their source(s) (e.g. CALINE4 \citep{benson1992}). Since these simulations are computationally rather expensive, such models provide predictions on a grid where each grid cell covers a large spatial area (e.g. CMAQ provides predictions on grid cells of size 12km x 12km). Epidemiologic studies utilizing these models determine the cell within which a subject residence or work place is located, and assign the corresponding pollutant exposure to the subject. Some recent studies have combined these two approaches together to develop data fusion models. In these models, data from both observed concentrations at limited sites and outputs from deterministic computer models are jointly used to provide predictions at unsampled locations at a fine spatial and temporal resolution \citep{sacks2014, turner2016}. 

Regardless of the sophistication of the method used to assign pollutant exposure to subjects using their residential or work locations, a fundamental challenge that is not addressed is the fact that subjects do not spend all of their time at home or at work. Humans move around during the day, and the patterns of movement likely depends on a number of factors such as time of day, day of week, season, employment status, etc. By not taking mobility into account, current exposure models may suffer from significant bias in their estimates. 

The goal of this study is to determine the distribution of exposure bias to an air pollutant (\ozonens) when mobility is not taken into account, and to identify groups of individuals for which this bias is large. To achieve this goal, we require information on patterns of human mobility. Mobility is well-captured by cell phone data; however, most available cell-phone based mobility studies either require users to install an ``app'' to capture their location \citep{bayir2009} or rely on other, more opportunistic approaches \citep{calabrese2011}. These methods suffer from not sufficiently sampling at the population level and are likely not robust enough to generalize. 

Analyses have been performed using cell phone infrastructure based on Call Detail Records (CDRs), which are generated by applications or phone calls on cellular devices \citep{becker2013, lu2017, thuillier2018, marques2018}. These records are generated primarily for billing purposes and include information including location (triangulated via towers), data usage, etc. These records give precise location and duration information about the device while it is in use; however, they do not provide information at times when the device is not in use, and therefore have substantial time windows with no information. Analyzing mobility using such data likely leads to biased results, as the data are not representative of mobility for individuals themselves, and for the entire population. 

The number of studies that explicitly account for human mobility when assigning exposure to air pollutants is fairly small. \citet{nyhan2018} used CDR data to incorporate a work and home location into pollution exposure assignment, while \citet{ozkaynak2009} incorporated human movement indirectly through the Environmental Protection Agency's (EPA) Consolidated Human Activity Database. \citet{warren2017} incorporated longer-term movement, such as relocation to a different state, to analyze susceptibility to environmental pollution during pregnancy.

This study uses Advanced Wireless Service (AWS) data, which provides the most passive and complete data on population-level mobility. The data consists of cell devices that are turned on, are connected to towers, and that are serviced by at least two towers (discussed in Section 2). Very few studies that quantify population-level human mobility at the census tract-level use cell data at the infrastructure level \citep{becker2013} and, to our knowledge, this is the first study that integrates pollution information  with human mobility for an accurate exposure study at this scale. Our approach uses cell-tower level data to determine patterns of human mobility. We assign exposure to \ozone concentration to mobile devices, which are used as proxies for individuals, in the US state of Connecticut (CT) during a week in summer, 2016. Based on individuals' mobility behavior and the tower area they are connected to most between the hours of 8:00 PM and 6:30 AM, which we term the {\em night-time local area}, we are able to assess the exposure assignment bias. 

Knowledge of the distribution of \ozone exposure assignment bias will be extremely beneficial for environmental epidemiologists studying the effects of \ozone exposure on adverse health outcomes. Minimal exposure assignment bias due to a static pollutant concentration assignment would validate the results of past studies, while a significant exposure assignment bias would provide evidence for the need for human mobility to be explicitly accounted for in future epidemiologic studies. 

This paper proceeds as follows. In section 2, we provide details of the cell-phone telemetry data. In section 3, we provide details on the data and model used to provide predictions of \ozone concentrations at the cell tower sites. In section 4, we provide details on the distribution of \ozone exposure assignment bias, with concluding remarks in section 5.

\section{Quantifying Human Mobility with Cell-Tower Hand-offs}
\label{human-mobility-section}

\subsection{Cell-Tower Density in the State of Connecticut}
\begin{figure}[htbp!]
\centering
\includegraphics[width=0.8\textwidth]{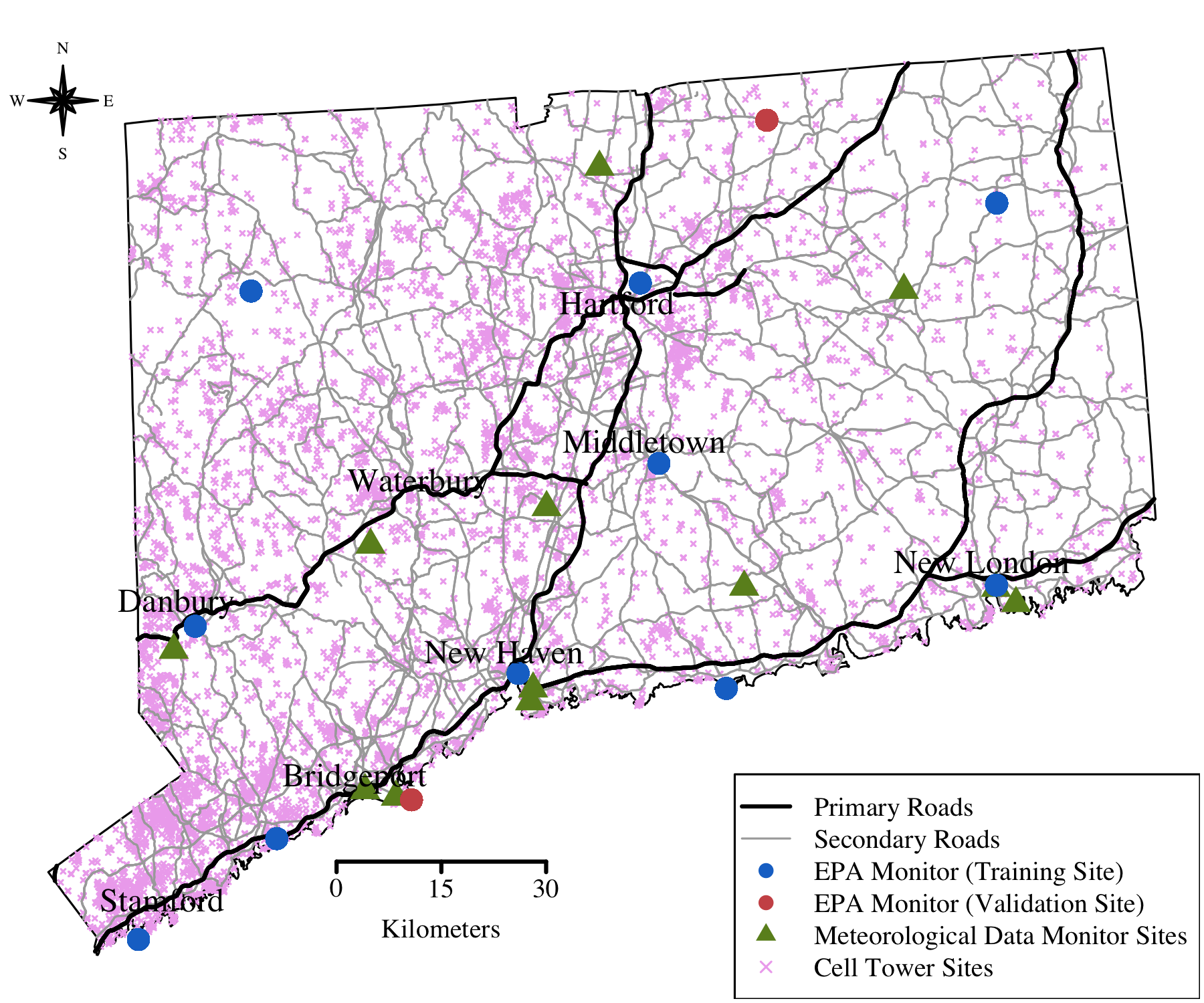}
\caption{The locations of primary and secondary roads; EPA and meteorological monitors; and and cell tower locations for the state of Connecticut (from OpenCelliD (https://opencellid.org))}
\label{monitoring_stations}
\end{figure}

Cellular phone towers coordinate data movement between cell phones and other cell-connected devices providing both phone calls and internet connectivity. Interconnected towers form a network capable of routing data traffic between cell devices or to internet-connected devices through a base station. U.S. cell phone networks facilitate the communication of hundreds of millions of devices, transferring vast amounts of data on any given day. This study is restricted to the state of Connecticut (CT). Figure \ref{monitoring_stations} shows the locations of approximately 10,000 cell towers in the state, along with the locations of the primary and secondary roads in the state, taken from the OpenCelliD website (https://opencellid.org). The data-traffic capacity of an individual tower is limited; to compensate for spatial areas with higher capacity demands, more towers may be erected in those areas. As a result, the spatial distribution of the towers matches the population distribution. This is seen in Figure \ref{monitoring_stations}, with most people residing in the southwest corner, the shoreline, and in metropolitan areas including Hartford, Danbury, Middletown, and Waterbury.

\begin{figure}[htbp!]
\centering
\includegraphics[width=0.9\textwidth]{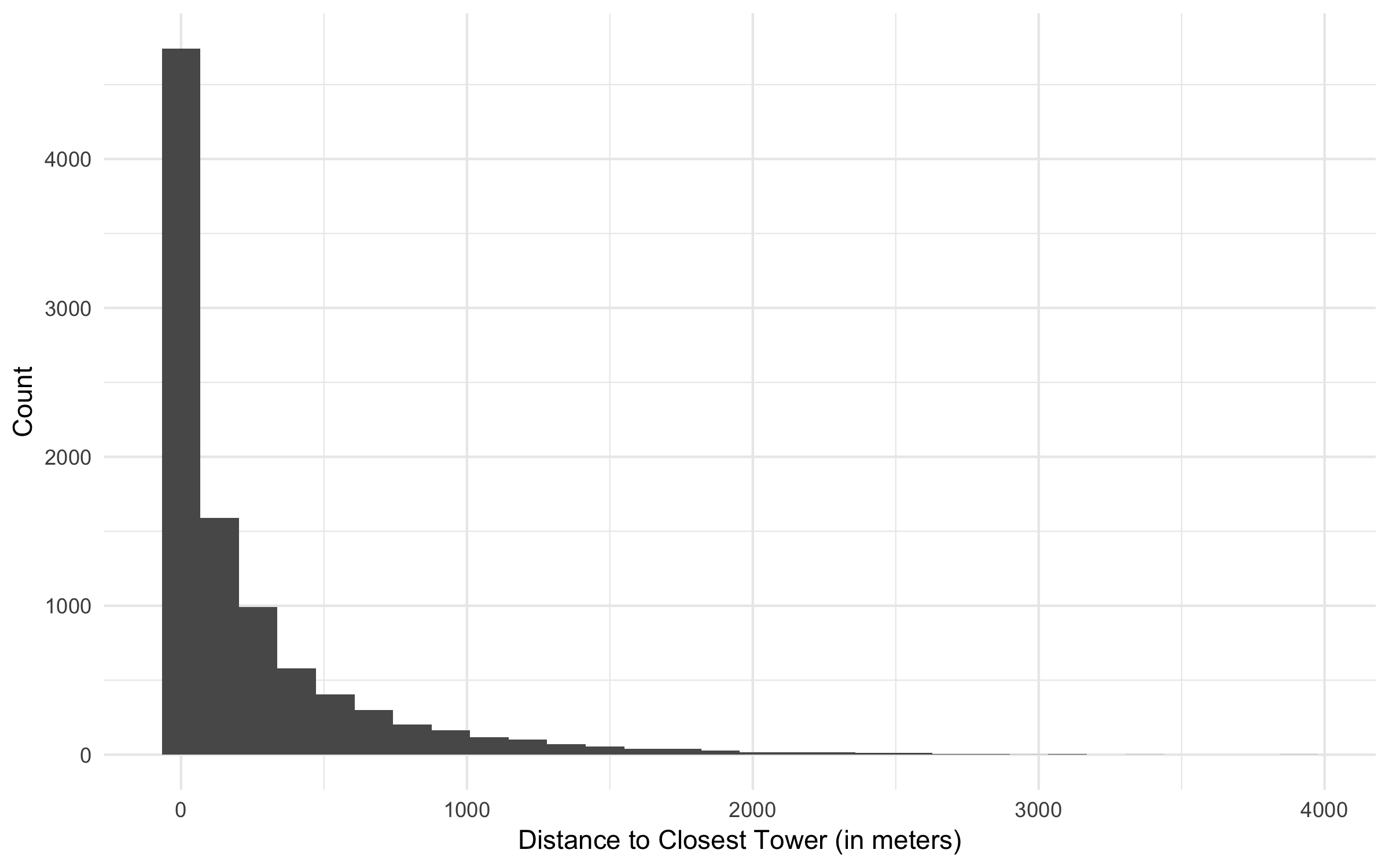}
\caption{Distribution of Haversine distance (in meters) to the nearest cell tower.}
\label{closest_tower_dist}
\end{figure}

Figure \ref{closest_tower_dist} shows the distribution of the distance to the closest tower from each tower. The distribution is roughly exponential. The lower quartile, median, and upper quartile are 2.7, 68.2, and 306 meters respectively. Many towers are within meters of other towers because multiple towers are sometimes built at a single site. This is especially true in highly populated areas, like southwestern Connecticut, where there are more constraints placed on where a tower can be placed as well as the fact that more towers are needed to compensate for higher traffic demands. Eastern Connecticut is generally less populated than the central and southwestern portions of the state, and the towers in these regions may be several kilometers away from the next closest one.

\subsection{Hand-off Trajectories in the Cell-Tower Network}

A single cellular device connects to the network through a tower. Devices generally connect to the closest tower, however, this can vary due to geographic features, which can occlude communication, and the amount of traffic being routed through the towers. As a phone moves through the network, it is ``handed-off'' between towers. A hand-off is characterized by a unique, anonymous device identifier, a date and time, and the location of {\em the tower} to which the device was handed. Data on these hand-offs may be analyzed to evaluate traffic load, tower placement, and connection integrity. Sequences of hand-offs in time for a single device, which we will refer to as {\em trajectories}, can be used to locate the device, and the humans that carry them, up to the resolution of the order 10's to 1000's of meters, depending on the local cell-tower density.

This study considers devices from one major U.S. carrier with at least one hand-off, where all hand-offs for the device during the period of July 18--24, 2016 were within the state of Connecticut. A total of approximately 50,000,000 hand-offs were recorded from approximately 400,000 unique devices. Given the sensitivity of these data, several steps were taken to ensure privacy of all individuals. First, personally  identifying  characteristics  are not included in these data. Data for the same device are linked using an anonymous unique identifier, rather than a telephone number, and the anonymization was performed by a party not involved in the data analysis.  No demographic data are linked to any cell phone user or used in this study. Second, all results are presented as aggregates.  That is, no individual anonymous identifier was singled out for the study.  By observing and reporting only on the aggregates, we protect the privacy of all individuals.

\begin{figure}[htbp!]
\centering
\includegraphics[width=0.8\textwidth]{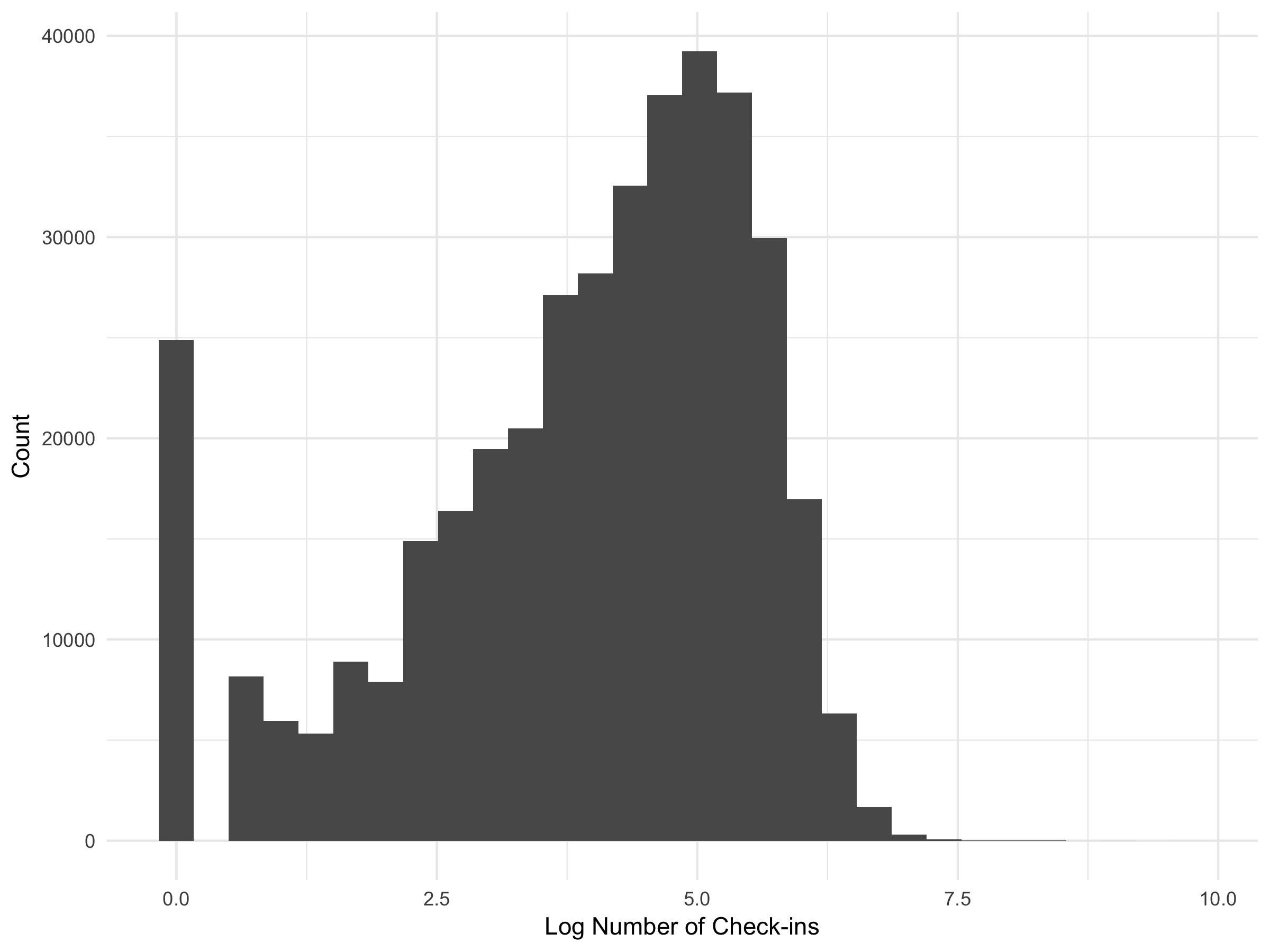}
\caption{A histogram of the log number of hand-offs.}
\label{imsi_hist}
\end{figure}

A histogram of the number of hand-offs per device is shown in Figure \ref{imsi_hist}. A total of approximately 25,000 devices had a single hand-off. The mode of the histogram where the log number of counts is greater than zero is centered around 5 and implies that most devices had approximately 150 hand-offs during the period considered.

\section{Description of Pollution Model} 
\label{pollution-model-section}

In order to determine hourly \ozone concentration exposure of users, we fitted a stationary Gaussian spatio-temporal model \citep{cressie2011statistics} to observed hourly O$_3$ concentrations from July 6 to August 5, 2016 for the state of Connecticut, and used it to predict \ozone concentrations at approximately 10,000 cell tower sites during the week of July 18--24, 2016. 

\subsection{Data}
%% \subsubsection{Ozone Data}

\begin{figure}[htbp!]
\centering
\includegraphics[width=0.9\textwidth]{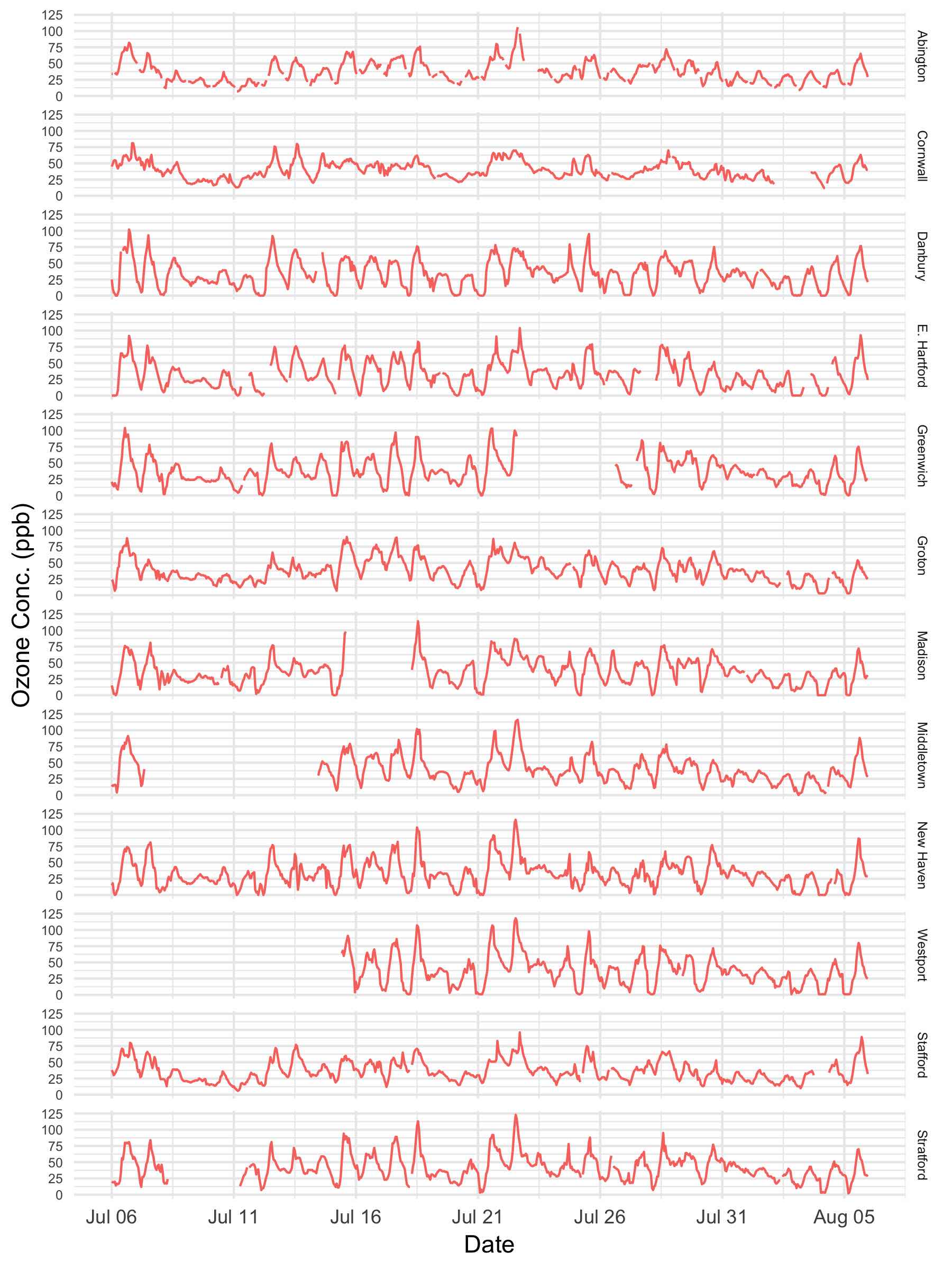}
\caption{Time-series plot of the observed \ozone concentrations (ppb) at the 12 monitoring sites in CT from July 6 to August 5, 2016.}
\label{ozone_ts}
\end{figure}

% PPB or ppb
We obtained data on observed hourly \ozone concentrations (in parts per billion - ppb) from the US Environmental Protection Agency (EPA) Air Quality System (AQS) Data Mart \citep{epaAQS} from July 6 to August 5, 2016 for the state of Connecticut. Figure \ref{monitoring_stations} shows the location of the 12 \ozone monitoring sites in CT; we randomly selected 10 sites for model fitting and two sites for model validation (Stafford and Stratford). At each site, \ozone concentration data were available for 744 time points. Figure \ref{ozone_ts} shows the observed hourly \ozone concentrations at the 12 monitoring sites. The two validation sites are displayed in the bottom two panels. The mean hourly \ozone concentration over the 31 days across the 10 training sites was 35.9 ppb with a standard deviation of 19.4 ppb, while the mean hourly concentration at the two validation sites was 39.0 ppb with a standard deviation of 17.9 ppb. 

% median = 34; 25%= 23; 75%=47; min=0; max=118

%%\subsubsection{Covariate Data}
In addition to data on the observed concentrations of \ozonens, we collected data on traffic and meteorological factors that could help explain the spatio-temporal variation observed in \ozone concentrations. Hourly temperature (in degrees Celsius) and wind speed (in meters per second) data were obtained from the National Oceanic and Atmospheric Administration (Quality Controlled Local Climatological Data) \citep{noaa}. Data were available from 12 weather stations in Connecticut (see Figure \ref{monitoring_stations}). For each \ozone monitoring site, we assigned hourly temperature and wind speed based on the measurements recorded at the closest weather station (see supplementary material for discussion on this choice). For hours with missing data, the last available hour for which data were available was used. Ozone concentrations tend to be lower in urban areas as compared to their surrounding rural areas - a phenomenon known as ozone urban decrement \citep{munir2012}. Using data on primary and secondary roads network in CT, available from the US Census Bureau \citep{tiger}, we calculated the minimum distance to primary and secondary roads (in meters). These variables served as proxies for population density in a neighborhood of the \ozone monitoring sites to account for ozone urban decrement. 

\subsection{Model}

As described by Cressie and Wikle (2011), a hierarchical spatio-temporal Gaussian Process (GP) model can be specified in three stages: the observed data stage, the true underlying process stage, and the parameter stage. We can specify distributions for the data, process, and parameters for each stage. The \texttt{R} package \texttt{spTimer} by \citet{bakar2015sptimer} can be used to efficiently fit GP models. Following the formulation of \citet{bakar2015sptimer}, let $Z(\s_i, t)$ denote the observed \ozone concentration at location $\s_i$, $i=1, \ldots, 10$, and time point $t, ~t=1, \ldots, 744$, and $Y(\s_i, t)$ denote the true underlying \ozone concentration at location $\s_i$ at time $t$. Let  $\mathbf{Z}_t = \left(Z(\s_1, t), \ldots, Z(\s_{10}, t)\right)'$ and $\mathbf{Y}_t = \left(Y(\s_1, t), \ldots, Y(\s_{10}, t)\right)'$ be the vectors of observed and true underlying \ozone concentrations, respectively, at time $t$. We define the nugget effect, or the pure error term, as $\boldsymbol{\epsilon}_t = \left(\epsilon(\s_1, t), \ldots, \epsilon(\s_{10}, t) \right)'$ to be independent (across space and time) and Normally distributed $\mathcal{N} \left(\mathbf{0}, ~\sigma^2_{\epsilon} \mathbf{I}_{10} \right)$, where $\sigma^2_\epsilon$ is the unknown pure error variance, and $\mathbf{I}_{10}$ is the 10 x 10 identity matrix.  Similarly, we denote the spatial random effects with independent replicates in time as $\boldsymbol{\eta}_t =\left(\eta(\s_1, t), \ldots, \eta(\s_{10}, t) \right)'$, assumed to be Normally distributed $\mathcal{N} \left(\mathbf{0}, ~\sigma^2_{\eta} \mathit{S}_{\eta} \right)$ and  independent of $\boldsymbol{\epsilon}_t$, where $\sigma^2_{\eta}$ is the site invariant spatial variance (also called the sill) and $\mathit{S}_\eta$ is the spatial correlation matrix. We assume that the spatial correlation can be modeled by the exponential function, so that the covariance between two locations $\s_i$ and $\s_j$ is a function of the the Euclidean distance $d_{ij}$ between the sites, i.e., $Cov \left(\eta(\s_i, t), \eta(\s_j, t)\right) = \sigma^2_{\eta}\cdot \mbox{exp}^{-\phi d_{ij}}$. Further, let $\mathbf{X}_t$ be a 10 x 5 matrix of covariates (including a column of 1s for the intercept) and $\boldsymbol{\beta} = \left(\beta_0, \ldots, \beta_4 \right)$ denote the 5 x 1 vector of unknown regression coefficients. 

We can then specify the hierarchical GP model by:

\begin{eqnarray}
\mathbf{Z}_t & = & \mathbf{Y}_t + \boldsymbol{\epsilon}_t \\
\mathbf{Y}_t & = & \mathbf{X}_t \boldsymbol{\beta} + \boldsymbol{\eta}_t
\end{eqnarray}

Details on fitting the model and obtaining parameter estimates using the package \texttt{spTimer}, including the full conditional distributions of the parameters, are given in \citet{bakar2015sptimer}. We used the default recommendations from the package \texttt{spTimer} for the initial values and the values of the hyper-parameters for the prior distributions. Specifically, we assigned flat Normal priors centered at 0 with large variances ($10^{10}$) for the regression coefficients; and flat Inverse-Gamma priors for the variance components $\sigma^2_{\epsilon}$ and $\sigma^2_{\eta}$. The value for the spatial decay parameter ($\phi$) in the spatial covariance matrix was fixed at 3/$d_{max}$, where $d_{max}$ is the maximum distance between the ozone monitor sites. Various alternative fixed values were tested, as well as estimating $\phi$ using a Uniform prior distribution. However, the predictive performance was best for the model that used the default fixed value of $\phi$ (0.0186). While the model is specified in the higher language \texttt{R}, the package \texttt{spTimer} performs calculations in the lower level language \texttt{C} for much faster computation. As per default, the Markov chain Monte Carlo was run for 4,000 further iterations after discarding the first 1,000 as burn-in. MCMC diagnostics were performed using package \texttt{coda} \citep{plummer2006coda}; all MCMC chains had converged during the burn-in, and auto-correlation plots displayed independence between iterations. The residual plot also did not show any departures from normality. 

\begin{table}
\caption{Posterior mean, median, standard deviation, and 95\% credible interval for the regression and covariance parameters.}
\begin{center}
\begin{tabular}{l r r r c }
\hline 
\hline
 Parameter &            Mean & Median &   SD & 95\% CI \\
\hline
 $\beta_0$ (Intercept)  &  10.7 & 10.7 &1.27   &(8.27, 13.2) \\
 $\beta_1$ (Temperature) &   1.02 &  1.02 &0.05   & (0.92,  1.13) \\
 $\beta_2$ (Wind Speed) &  -0.41 & -0.41 &0.10  &(-0.60, -0.23) \\
 $\beta_3$ (Dist. to Prim. Rd)&   0.0002 &  0.0002 &0.0000  & (0.0001, 0.0002) \\
 $\beta_4$ (Dist. to Sec. Rd) &   0.0005 &  0.0005 &0.0001  & (0.0003, 0.0007) \\ 
 \vspace{-9pt} \\
 \hline
 \vspace{-9pt} \\
 $\sigma^2_\epsilon$ (Nugget)   &  13.2 & 13.2 &0.79  & (11.8, 14.8) \\
 $\sigma^2_\eta$ (Sill)   & 174.9 &174.9 &2.87 & (169.3, 180.4) \\
 \hline
\end{tabular}
\label{tab:par_estimates}
\end{center}
\end{table}

Estimated parameters from the model are given in Table \ref{tab:par_estimates}. As expected, higher temperature is associated with increased \ozone concentrations while increased wind speed is associated with lower \ozone concentration. Minimum distance to primary and to secondary roads both have positive slopes, indicating the expected ozone urban decrement. All coefficients are statistically significant since none of the 95\% credible intervals contain 0. 

\subsection{Prediction}

Once the GP model has been fitted and posterior distributions for the unknown model parameters have been obtained, spatial prediction at a new location $\s_0$ (and temporal prediction at a future time point $t'$) can be obtained using the posterior predictive distribution for $Z(\s_0, t')$. The function \texttt{predict} in the package \texttt{spTimer} provides spatio-temporal predictions (see \citet{bakar2015sptimer} for technical details). We predicted hourly \ozone concentrations for the 31 day period from July 6 to August 5, 2016 at the two validation EPA \ozone monitoring sites, as well as at the approximately 10,000 cell tower sites.

\begin{figure}[htbp!]
\centering
\includegraphics[width=0.9\textwidth]{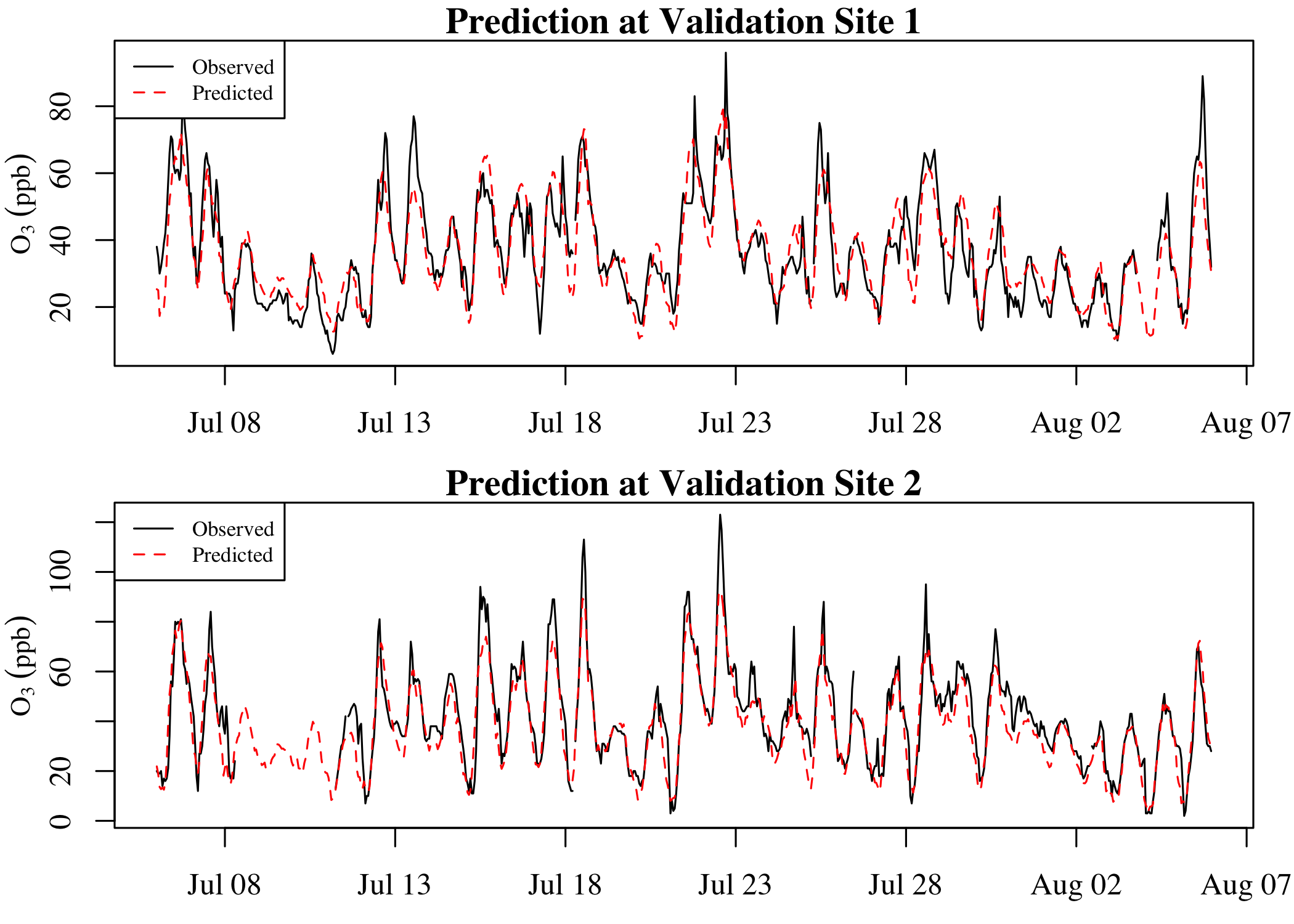}
\caption{Comparison of the observed (black solid) and predicted (red dashed) \ozone concentrations (ppb) at the two validation sites.}
\label{fig_validation}
\end{figure}

\begin{figure}[htbp!]
\centering
\includegraphics[width=\textwidth]{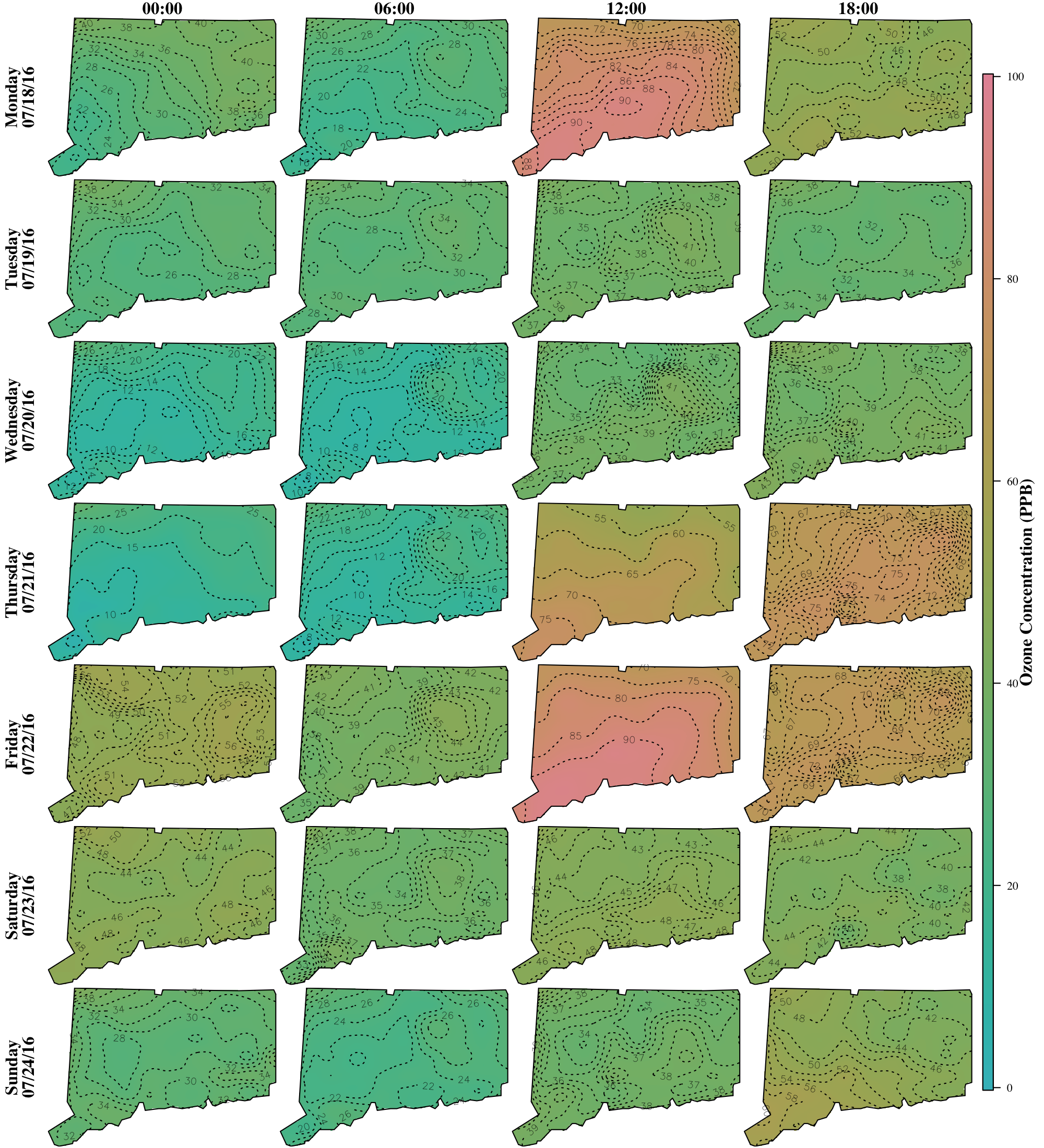}
\caption{Prediction of ozone concentrations at midnight, 6 AM, noon, and 6 PM for each of the seven days in the study.}
\label{fig:CTpred}
\end{figure}

Figure \ref{fig_validation} shows a time-series comparison of the observed and predicted hourly \ozone concentrations at the two validation sites. At both sites, the predicted concentrations are very similar to the observed concentrations. The root mean square error at the validation sites is 7.68, while the relative bias is -0.0461 and the relative mean separation is 0.1811. The small values for the relative bias and relative mean separation suggest a fairly good model fit. Figure \ref{fig:CTpred} shows a prediction map for hourly \ozone concentrations across CT at midnight, 6 am, noon, and 6 pm on seven consecutive days from July 18-24, 2016. It shows that the distribution of \ozone concentration is somewhat homogeneous across space, but changes considerably throughout the day. There also appears to be a concentration gradient in the southwest to northeast direction, particularly during the day-time hours, with higher concentrations in the southwestern part of the state. 

% Table \ref{tab:validation} provides validation metrics for the prediction at these sites.

%\begin{table}
%\caption{Prediction comparison at validation sites: Mean Square Error (MSE), Root Mean Square Error (RMSE), Mean Absolute Error (MAE), Mean Absolute Percentage Error (MAPE), Bias (BIAS), Relative Bias (rBIAS), and Relative Mean Separation (rMSEP).}
%\begin{center}
%\begin{tabular}{r r r r r r r }
%\hline 
%\hline
%   MSE &    RMSE  &   MAE &   MAPE &   BIAS &  rBIAS &  rMSEP  \\
%\hline
% 58.9467 & 7.6777 & 5.8489 & 17.3364 & -1.7999 & -0.0461 & 0.1811 \\
%\hline
%\end{tabular}
%\label{tab:validation}
%\end{center}
%\end{table}

\section{Exposure Distributions}

As described in section \ref{human-mobility-section}, a device hand-off is characterized by a unique, anonymous device identifier, a date and time, and the location of the tower to which the device was handed. From sequences of hand-offs, sorted by date and time, the duration over which the device was checked into the tower was calculated. In addition, a new feature was calculated per device over the study duration - the tower that a device was checked into for the longest duration during the hours of 8:00 PM and 6:30 AM. We note that this new variable is correlated with the night-time local area for working individuals not participating in shift work. However, they are distinct in that, due to de-identification we don't know the residence or occupational characteristics of the devices' owners.

For each tower location the hourly \ozone concentration was estimated as described in section \ref{pollution-model-section}. The pollution estimate can be merged with the mobility data using a join over the date, time, and cell-tower locations. The individual-level exposure estimate is then calculated as the amount of pollution at the tower location multiplied by the duration at that tower's catchment area. 

As per the standards set by the EPA, we calculated the average 8-hour maximum exposure for each individual for each of the seven days by calculating the hourly average ozone concentration for each 8-hour window during the day, and selecting the maximum of these hourly averages for each day.

\subsection{Distribution of the Difference in the Average 8-Hour Max Exposure}

\begin{figure}[htbp!]
\centering
\includegraphics[width=\textwidth]{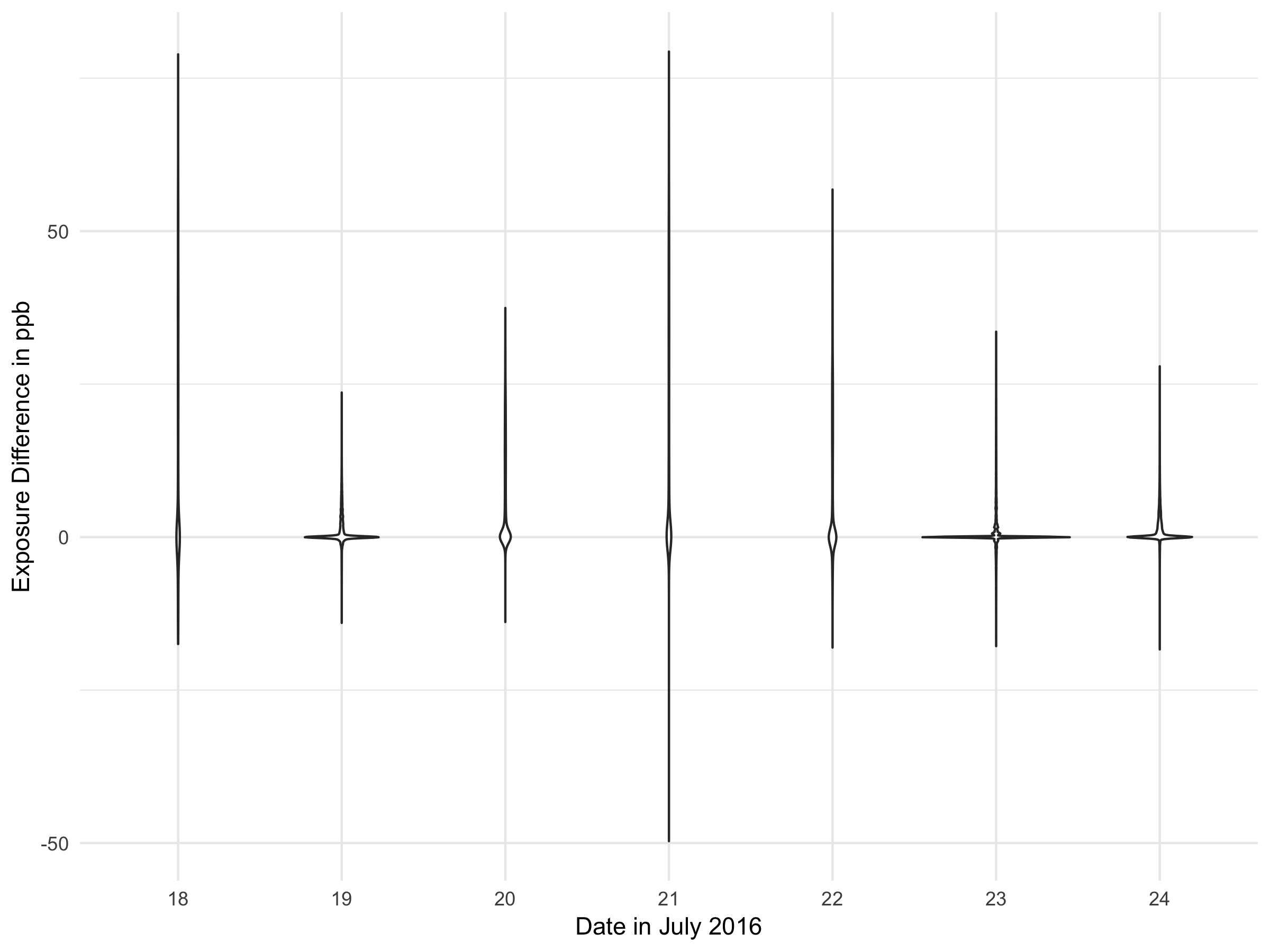}
\caption{Violin plot showing the distribution of average 8-hour max ozone exposure difference (dynamic scenario exposure minus the static scenario exposure) for each day of the study window.}
\label{fig_diff_prop}
\end{figure}

To assess the distribution of \ozone exposure assignment bias when mobility is not taken into account, we calculated the average 8-hour max exposure for each device per day under two scenarios: (a) using their trajectories to determine the tower catchment area visited as they moved throughout the day (which we will refer to as the dynamic scenario); and (b) assuming that the individuals spent the entire day at their night-time local area (which we will refer to as the static scenario), similar to what is typically done in epidemiologic studies (see Figure \ref{fig_8_hour_max} for a violin plot of these distributions). The exposure assignment bias for each device per day was then calculated as the difference between the dynamic and static average 8-hour max exposure assignment for each day. The result is shown in Figure \ref{fig_diff_prop}. (A similar plot showing the distribution of average hourly difference between the dynamic and static scenarios based on a 24-hour cumulative exposure - instead of the average 8-hour max exposure - is given in Figure \ref{fig:hourly_exp_diff_cum}, while Figure \ref{fig:hourly_exp_dynamic_static} shows the distribution of hourly exposure assignment for the dynamic and static scenarios). All of the estimated differences were within 80 ppb per hour. As a reference, the US EPA ozone air quality standard is set at 70 ppb (averaged over an 8-hour period); concentrations consistently exceeding this level are considered harmful to human health and welfare \citep{epaOzoneStd}. Since the mean and medians are close to zero, this result {\em validates} many of the current studies of ozone exposure for a large cross-section of the population, i.e., the exposure assignment bias due to not taking mobility into account is not too large. However, we observe that the distribution of the differences have heavy tails, and that the upper tails are longer than the lower tails. This suggests that the static scenario underestimates the true concentration more frequently than overestimating it. Additionally, it should be noted that differences are correlated for an individual device. That is, devices with large differences between the models at a given hour tend to have large differences at other hours. This implies that there are distinct subpopulations for which static scenarios are biased. This is further explored in subsequent sections.

\subsection{Weekend vs. Weekday Daily Exposure, Accounting for Mobility}

%For each of the trajectories, the exposure was calculated, per day, by summing the hourly exposure of each device. The distributions are shown in Figure \ref{exposure_density}, by hour, for both weekdays and weekend. Due to the size of the data, the plots are based on 10,000 sampled weekend exposures and an additional 10,000 weekday samples. The plots show that the minimum exposure is higher on the weekend when compared to the weekday, they show that the total range of exposure is greater on weekdays compared to weekend, and they show that the maximum exposure is higher on the weekdays. It should be noted that these results are likely driven by the variation in ozone concentrations seen on the weekday when compared to the weekend, as opposed to the device mobility.

\begin{figure}[htbp!]
\centering
\includegraphics[width=\textwidth]{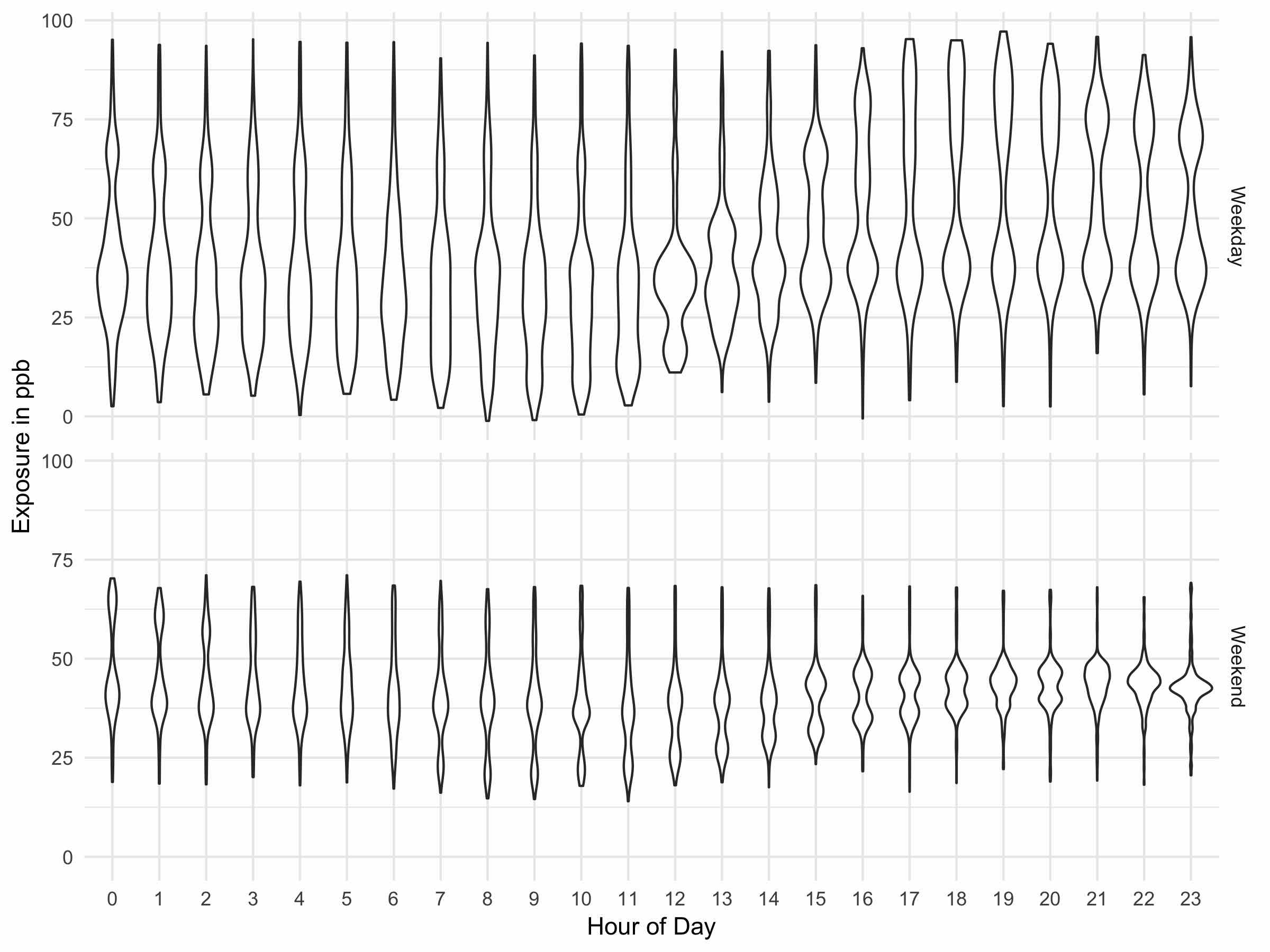}
\caption{Violin plots of the exposure distribution by hour of day for weekdays and weekends (using the dynamic scenario).}
\label{exposure_density}
\end{figure}

Figure \ref{exposure_density} shows the distribution of average hourly ozone exposure for individuals for each hour of the day for weekdays and weekends, taking mobility into account. Due to the size of the data, the plots are based on a random sample of 10,000 devices. The plots show that the minimum \ozone exposure is higher on the weekend (for every hour) as compared to the weekday, while the maximum exposure is higher on the weekdays. They also show that the range of average hourly \ozone exposure is greater on weekdays as compared to weekends. It should be noted that these results are likely primarily driven by the variation in \ozone concentrations seen on the weekday when compared to the weekend (as seen in Figure \ref{fig:CTpred}), as opposed to the device mobility. The plots also reveal a generally unimodal exposure distribution during the late night/early morning hours, but a bimodal exposure distribution during the afternoon and evening hours. This again probably reflects the smaller spatial variation in \ozone concentrations during the late night and early morning hours, and greater spatial variation during the afternoon and evening hours. (Figure \ref{cum_exposure_density} shows the distribution of the cumulative exposure by hour for weekdays and weekends, while Figure \ref{cum_exposure_density_day} shows the distribution of the cumulative exposure at the end of the day for weekdays and weekends.)

\subsection{Areas of Poorly-Predicted Exposure when Mobility is not Taken into Account}

While the bias from not taking mobility into account is close to zero for most individuals, for some it is up to 80 ppb/hour. While this bias may be negligible for a given day, cumulative exposure difference may result in very different health-outcomes than what is predicted by models not taking mobility into account. In particular, individuals with an exposure assignment bias of around 80 ppb/hour likely experience this assignment bias for 8-10 hours per day on multiple days, resulting in vastly different cumulative exposure assignment bias. 

To understand which class of individuals are under-served by existing models, device trajectories where the difference between exposure assessed using the dynamic versus static scenarios was in the highest and lowest  1\%, corresponding to individuals with higher and lower exposure with respect to the static scenario, were extracted and a contour plot was created showing their night-time local areas.

\begin{figure}[htbp!]
    \centering
    \begin{subfigure}[t]{0.7\textwidth}
        \centering
        \includegraphics[width=\textwidth]{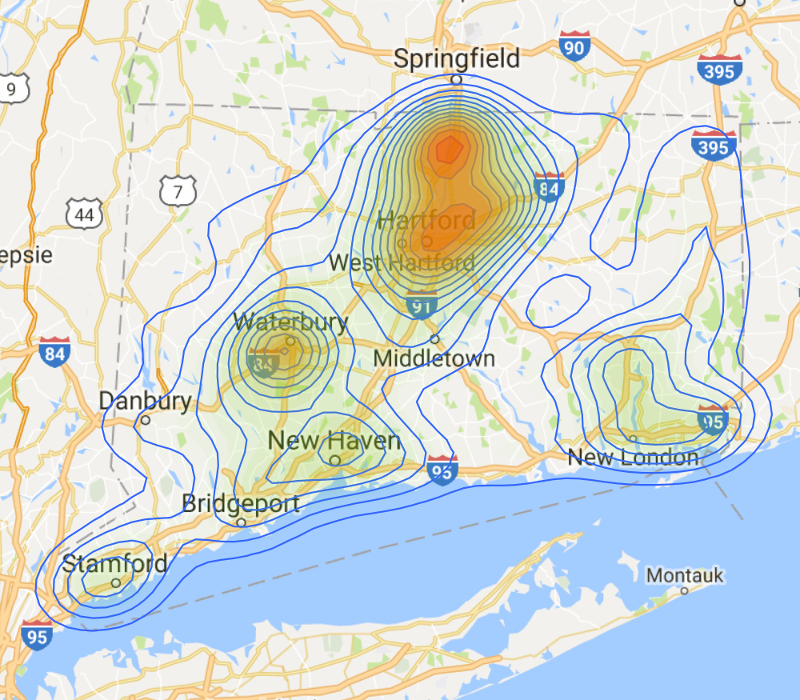}
        \caption{~}
    \end{subfigure}%
    
    \begin{subfigure}[t]{0.7\textwidth}
        \centering
        \includegraphics[width=\textwidth]{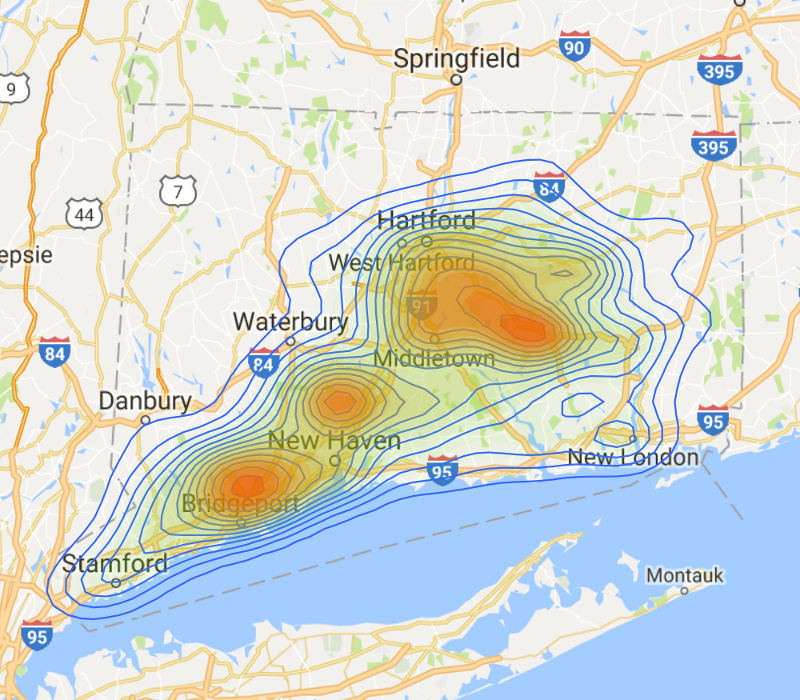}
        \caption{~}
    \end{subfigure}
    \caption{Night-time local-areas of individuals with (a) higher exposure in the dynamic scenario compared to the static scenario (top 1\%), and (b) lower exposure in the dynamic scenario compared to the static scenario (bottom 1\%)}
    \label{more_less_exposure}
\end{figure}

The night-time local area of individuals receiving more exposure when mobility is taken into account is shown in Figure \ref{more_less_exposure}a. The difference in exposure assessed based on the dynamic and static scenarios is generally due to individuals' mobility during day-time commuting hours. \ozone predictions at different times of the day given in Figure \ref{fig:CTpred} show that during the afternoon hours, \ozone concentrations are generally higher in the southwestern part of Connecticut (likely due to higher concentrations in New York City area), and decrease in a northeasterly direction. Comparing Figure \ref{more_less_exposure} to \ozone predictions shown in Figure \ref{fig:CTpred}, we observe that many of the individuals identified as receiving a considerably larger exposure when taking mobility into account are likely suburban residents commuting into the associated urban areas (in a western or southern direction) where ozone concentrations are higher during the day-time hours. In particular, the cluster of individuals residing north and northeast of Hartford likely commute to Hartford during the day, while the cluster identified around Waterbury are likely residents that commute to Danbury during the day.

The night-time local area of individuals receiving less exposure when mobility is taken into account is shown in Figure \ref{more_less_exposure}b. There are three distinct clusters: one is southeast of Hartford, one is north of New Haven, and one is north of Bridgeport. Many of these individuals are likely suburban residents commuting into the associated urban areas (in an eastern or northern direction) where ozone concentrations are lower during the day-time hours. In particular, the clusters southeast of Hartford and north of New Haven likely represent individuals that commute to Hartford and Middletown during the day, while the cluster north of Bridgeport perhaps represents individuals commuting northeast to New Haven.

These results suggest that the direction of mobility during day-time hours is an important factor determining the adequacy of the static scenario in accurately assessing individual exposure to \ozone concentration. Movement along the pollutant concentration gradient naturally results in a higher difference between actual exposure and the exposure modeled assuming static behavior.

%\subsection{Morning vs. Evening Commuting}

%\subsection{Exposure by Home Census Tract}

%\subsection{Exposure by Work Census Tract}

%\subsection{Exposure Efficiency}

\section{Conclusions}

This paper integrates mobility data from cell-tower hand-offs with a pollution model to more realistically estimate ozone exposure during the period of July 18--24, 2016. These estimates were compared to those of a model where mobility data are not available, which is commonly seen in the literature. We show that the bias introduced by not taking mobility into account is minimal for the majority of individuals in the state of Connecticut, thereby validating many existing epidemiological studies examining the health impacts of exposure to \ozone on various outcomes. However, we also show that existing models do a poor job estimating exposure of individuals who routinely commute into and out of urban areas, particularly whose day-time movements follow the pollution concentration gradient, and, in these cases, mobility should be taken into account.

A key strength of our analysis lies in the use of mobility data on the individual level for a very large portion of the state's population, which allows us to more accurately capture real mobility patterns. However, there are a few limitations to this analysis as well. 

The \ozone exposure model was developed using ten \ozone monitoring sites for the entire state of Connecticut. Given the spatial distribution of these sites, it is difficult to capture the small scale spatial variability in \ozone concentrations. However, this may not be too big of a concern in this particular analysis since \ozone concentrations are known to be fairly homogeneous over short distances. 

Data on mobility in this study were captured using hand-offs at cell towers, and not using actual GPS coordinates. Therefore, the mobility trajectories provide an estimate of where cellular devices are at any given point, and not their exact location. However, most cell towers are within 1000 m of each other, and therefore, cell device areas are generally within a 500 m buffer. This approach has the added advantage of protecting individuals' privacy in that their exact location is never known.

Due to practical reasons, we restricted our analysis exclusively to users who had at least one tower check-in during the one week window of analysis and stayed within the boundaries of CT during that week. This potentially has two issues related to generalizability of our results. First, users might not represent the general US population. We are not too concerned about this issue since given the competitive telecommunications market, there does not appear to be any strong evidence suggesting that this cellular service provider attracts a particular niche of customers. Additionally, an internal study found that their subscribers were well-represented geographically. Second, individuals that spend the entire week within the state of CT do not include long distance commuters. Long-distance commuters would be expected to have a much larger difference in exposure assignment between the dynamic and static scenarios, which would potentially lead to longer tails in the exposure difference distribution. Third, this study does not capture individuals who do not move within the one week period. However, this population is likely small and inhabiting an indoor space for the duration where they are not exposed to outdoor pollutants.

To our knowledge, this is the first study that compares exposure to a pollutant concentration based on the assumption of static behavior versus dynamic mobility for a significant portion (between 30 and 45\%) of the domestic cell user population for an area while at the same time avoiding the bias associated with CDR records. While the results of this study strengthen our confidence in the findings of epidemiologic studies looking at the adverse impact of \ozone concentration on various health outcomes, our analysis can be extended in two directions: looking at other states, or even the entire US, instead of just CT and over longer time duration; and analyzing other air pollutants, such as nitrogen oxides or small particulate matter, which are known to disperse quickly over short distances \citep{baldwin2015}. For such pollutants, we expect there to be a significant difference in exposure assignment between the dynamic and static scenarios. However, modeling concentrations for pollutants that vary rapidly over short distances is challenging, as they require a  rather dense network of pollutant monitors, which is generally not available. While data on human mobility are available to us in near real-time, a bottleneck arises in accurate air pollutant modeling at fine spatial and temporal scales. Additionally, extending the analysis to other states or the entire US over longer time duration presents challenges of dealing with vast quantities of data for both the air pollution modeling part and for human mobility. However, with rapid advancements in methods dealing with analysis of Big Data, this may not be a big challenge in the future.

\section*{Acknowledgements}
The authors would like to thank Drs. Janneane Gent, Ted Holford, and Joshua Warren for their suggestions in improving the analysis, as well as the anonymous reviewers for their helpful suggestions.

%\section{Research Outlook}

%% The Appendices part is started with the command \appendix;
%% appendix sections are then done as normal sections
%% \appendix

%% \section{}
%% \label{}

%% References
%%
%% Following citation commands can be used in the body text:
%% Usage of \cite is as follows:
%%   \cite{key}          ==>>  [#]
%%   \cite[chap. 2]{key} ==>>  [#, chap. 2]
%%   \citet{key}         ==>>  Author [#]

%% References with bibTeX database:

%\section*{References}

%\biboptions{colon,round}
\bibliographystyle{apalike}
\bibliography{sample.bib}

\newpage
\resetlinenumber
\setcounter{page}{1}
\renewcommand\thefigure{S.\arabic{figure}}  
\setcounter{figure}{0} 

\section*{Supplementary Material}

\subsection*{Spatial Distribution of Meteorological variables}

\begin{figure}[htbp!]
\centering
\includegraphics[height=0.8\textheight]{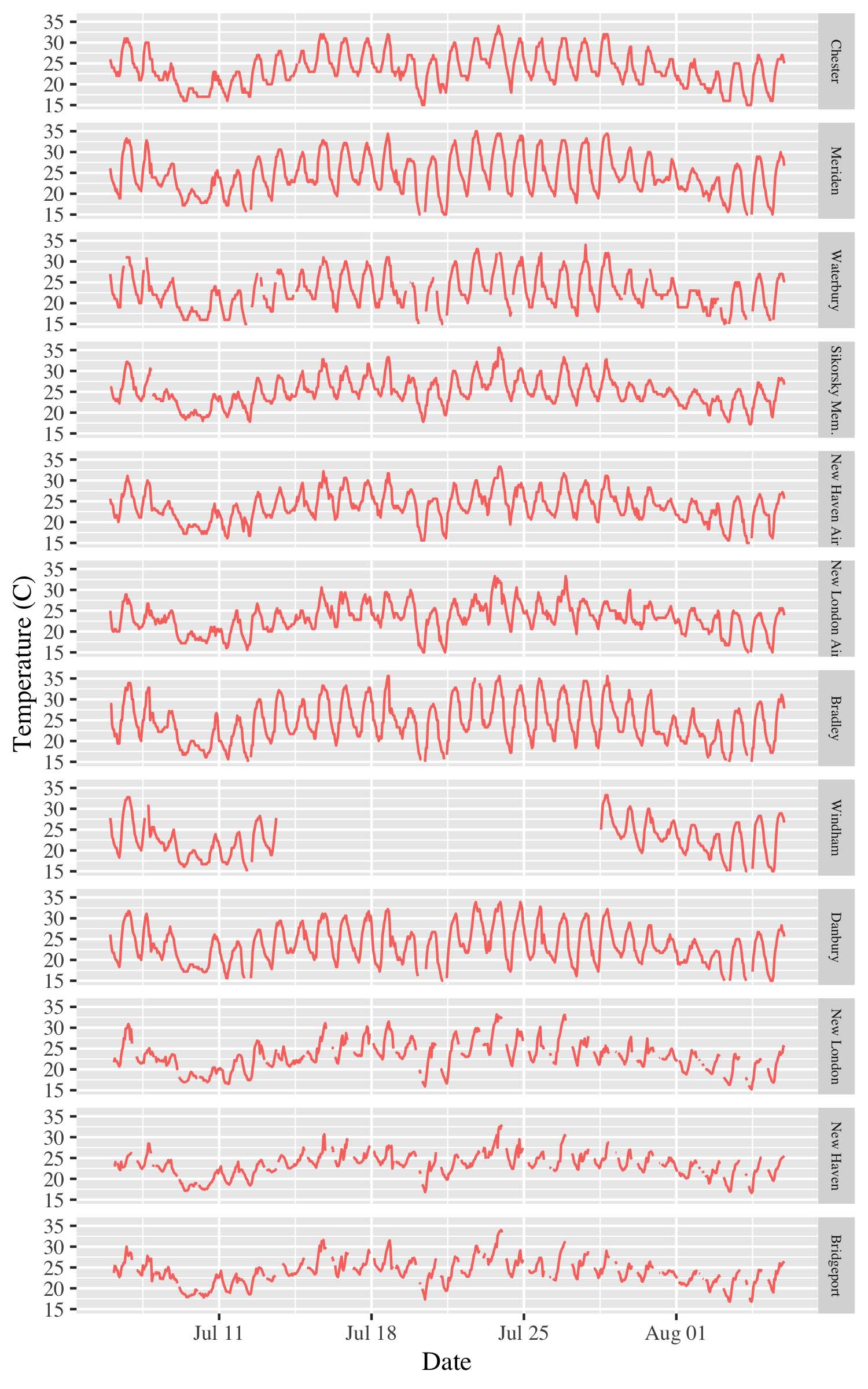}
\caption{Time-series plots of observed temperature (Celsius) at 12 meteorological stations during the study window.}
\label{fig_temp_all}
\end{figure}

\begin{figure}[htbp!]
\centering
\includegraphics[height=0.8\textheight]{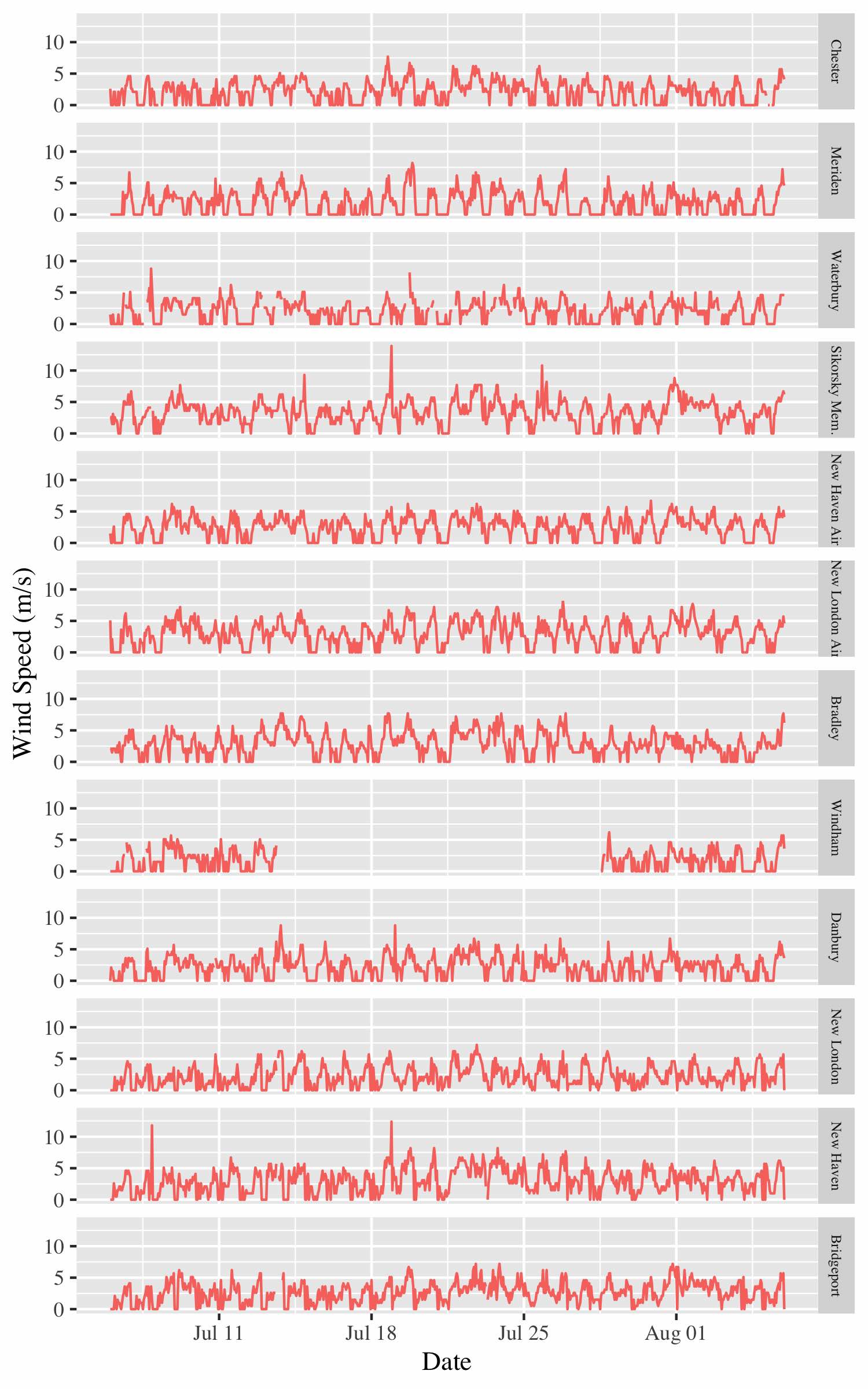}
\caption{Time-series plots of observed wind speed (m/s) at 12 meteorological stations during the study window.}
\label{fig_windspeed_all}
\end{figure}

\begin{figure}[htbp!]
\centering
\includegraphics[height=0.85\textheight]{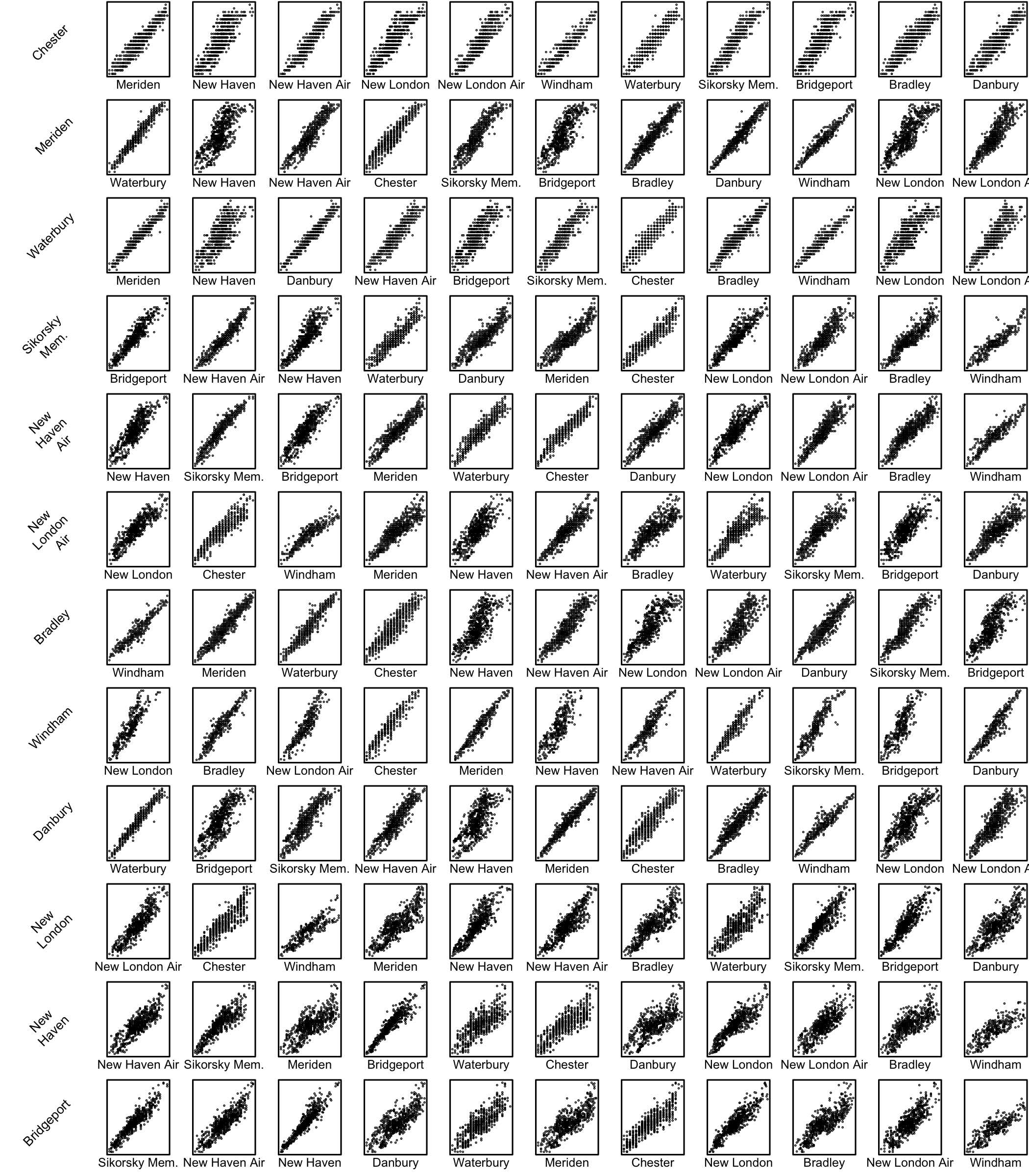}
\caption{Scatterplot matrix for temperature at the 12 meteorological stations. Each row corresponds to a different station. Within each row, the scatterplots are sorted by increasing distance to that particular station, with the left-most scatterplot corresponding to the \emph{closest} station.}
\label{fig_temp_corr}
\end{figure}

\begin{figure}[htbp!]
\centering
\includegraphics[height=0.85\textheight]{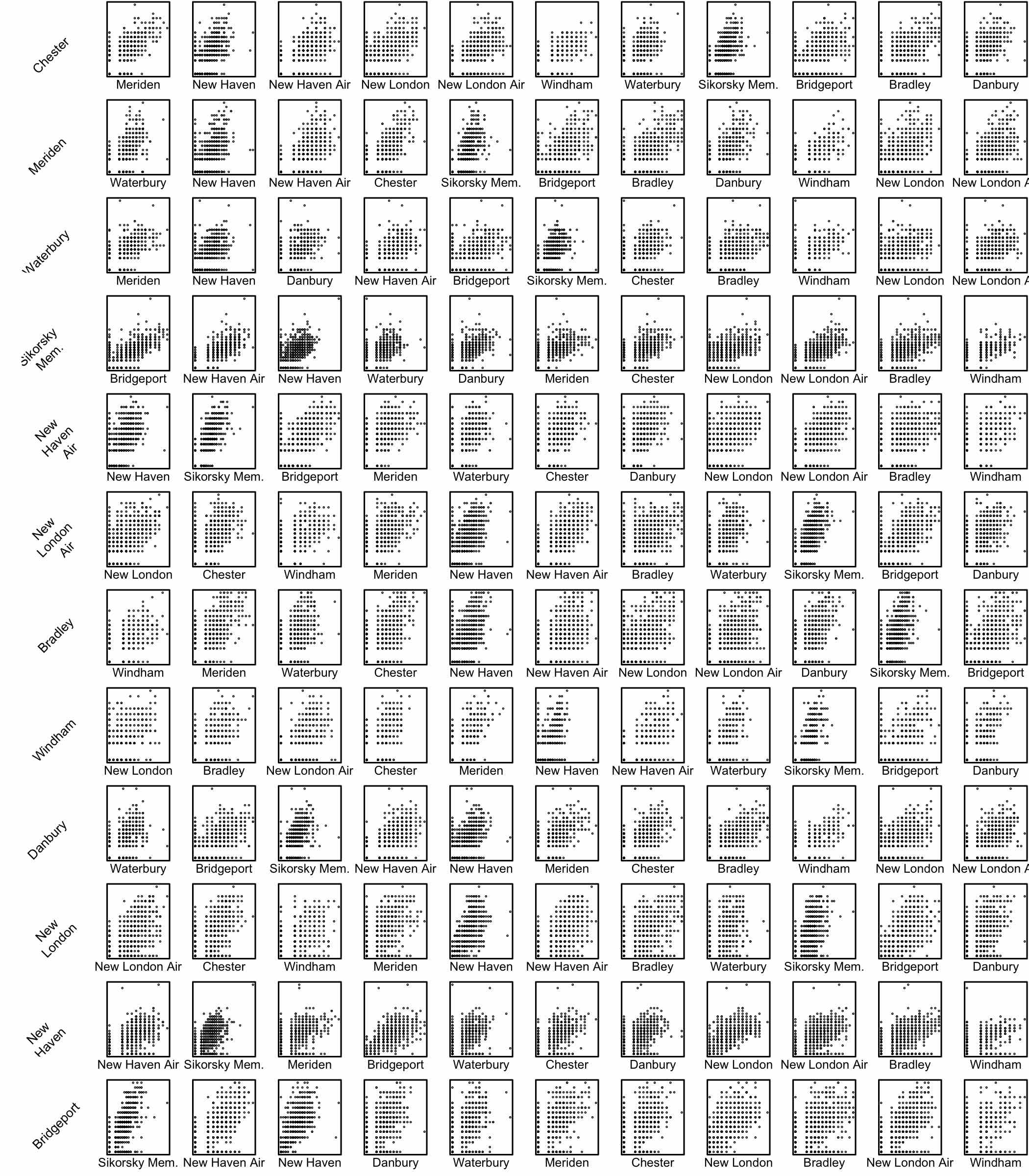}
\caption{Scatterplot matrix for wind speed at the 12 meteorological stations. Each row corresponds to a different station. Within each row, the scatterplots are sorted by increasing distance to that particular station, with the left-most scatterplot corresponding to the \emph{closest} station.}
\label{fig_windspeed_corr}
\end{figure}

\begin{figure}[htbp!]
\centering
\includegraphics[width=0.6\textwidth]{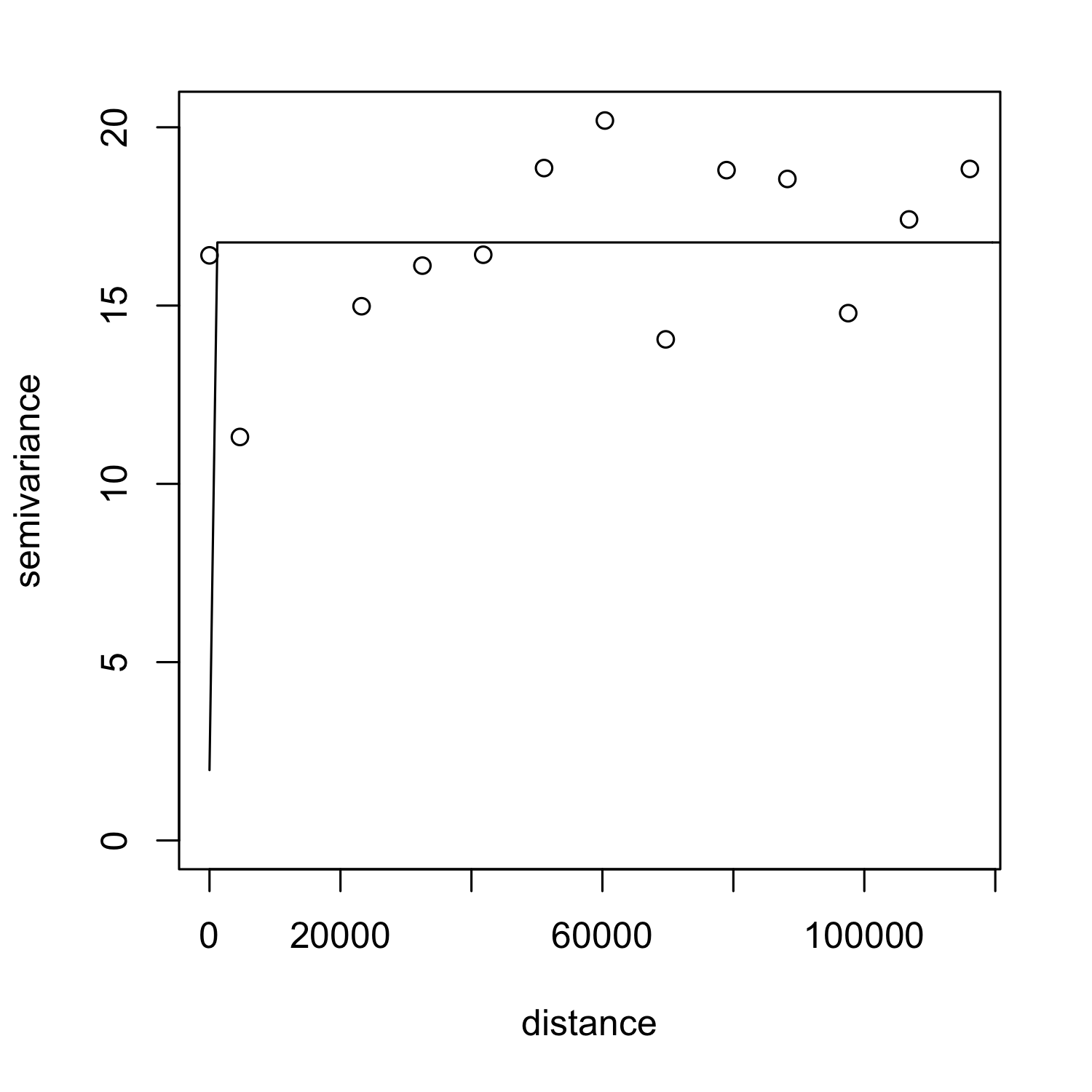}
\caption{Empirical and fitted semivariogram plot for temperature.}
\label{fig_temp_covariog}
\end{figure}

\begin{figure}[htbp!]
\centering
\includegraphics[width=0.6\textwidth]{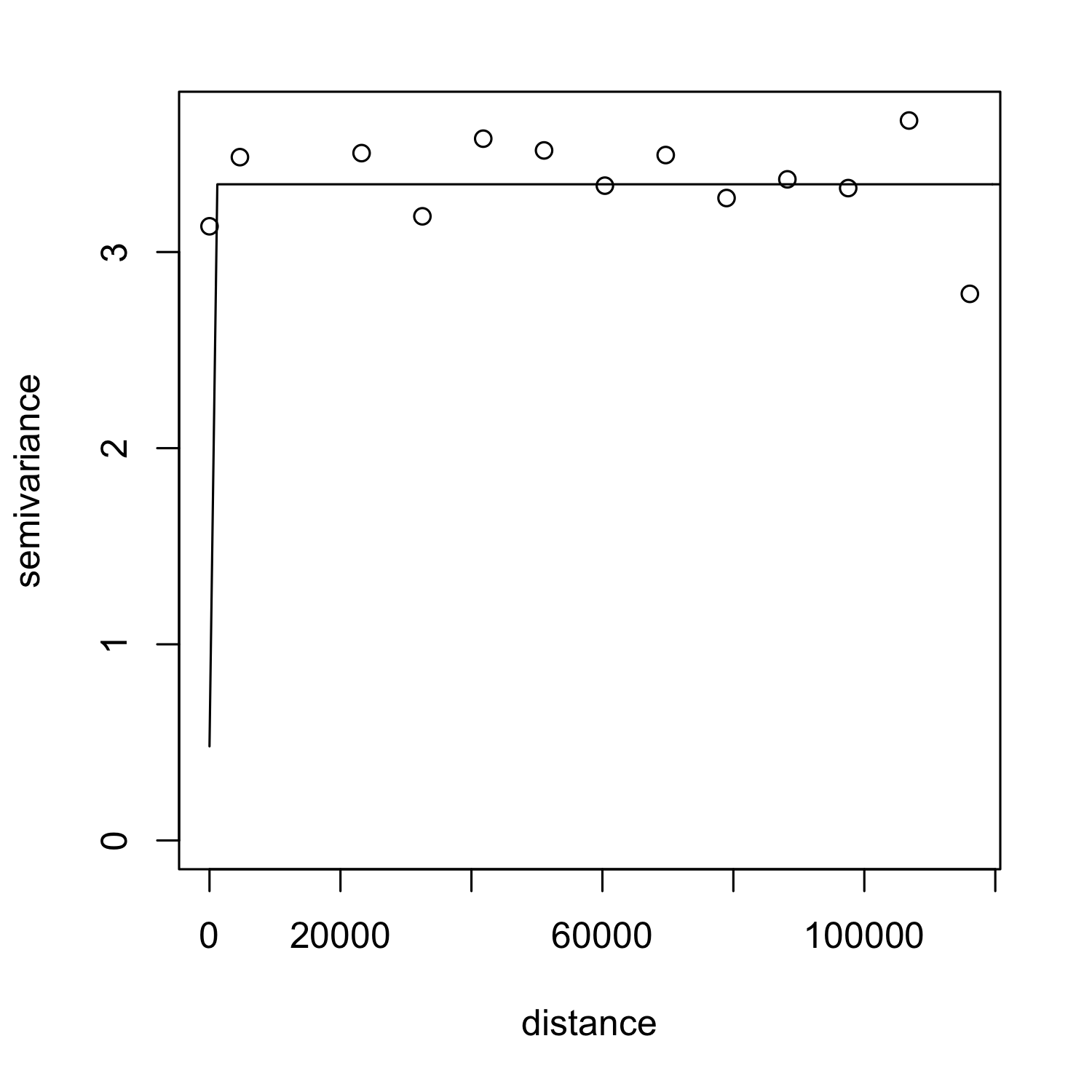}
\caption{Empirical and fitted semivariogram plot  for wind speed.}
\label{fig_windspeed_covariog}
\end{figure}

The following plots explore the spatial distribution of the meteorological variables (temperature and wind speed). In general, there was not a great deal of spatial variability in both temperature and windspeed over the study window.  Figures \ref{fig_temp_all} and \ref{fig_windspeed_all} show time series plots of monitored temperature and windspeed, respectively, for all 12 meteorological stations. As can be observed, the time-series follow fairly similar trends. Figures \ref{fig_temp_corr} and \ref{fig_windspeed_corr} show scatterplot matrices for temperature and wind speed, respectively, where the columns for each row are sorted by increasing distance. These plots also suggest little spatial variation in these two variables. Additionally, Figures \ref{fig_temp_covariog} and \ref{fig_windspeed_covariog} show the empirical and fitted semivariogram plots for temperature and wind speed, respectively. These plots are very flat.\\
  
  Based on these plots, and in an attempt to keep the modeling procedure as simple as possible while still giving fairly useful results, we decided to use the closest weather station to assign covariates values for the meteorological variables in our model.
\newpage

\subsection*{The Average 8-Hour Max Exposure Density}

\begin{figure}[htbp!]
\centering
\includegraphics[width=\textwidth]{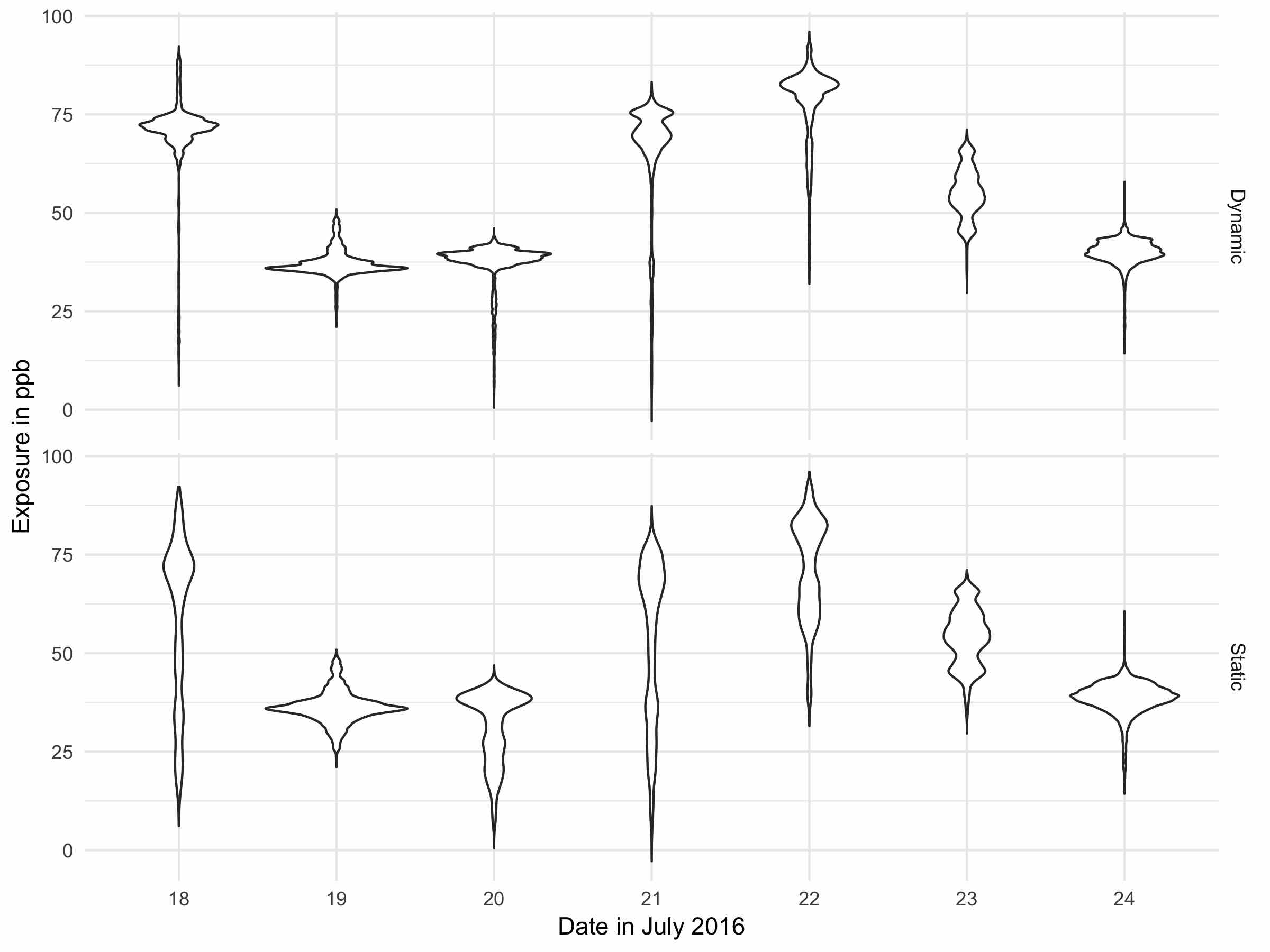}
\caption{The average 8-hour max distribution using the dynamic and the static scenario.}
\label{fig_8_hour_max}
\end{figure}

Figure \ref{fig_8_hour_max} shows the distribution of the daily average 8-hour max \ozone exposure for individuals for the week of July 18-24, 2016 for both the dynamic and static scenarios. (The distribution of the difference between these two models is given in Figure \ref{fig_diff_prop}). The distributions for each day generally look similar; however, the static scenario appears to have heavier lower tails as compared to the dynamic scenario, particularly on the weekdays. 

\pagebreak

\subsection*{The Hourly Exposure Difference Between Dynamic and Static Scenarios}

\begin{figure}[htbp!]
\centering
\includegraphics[width=0.7\textwidth]{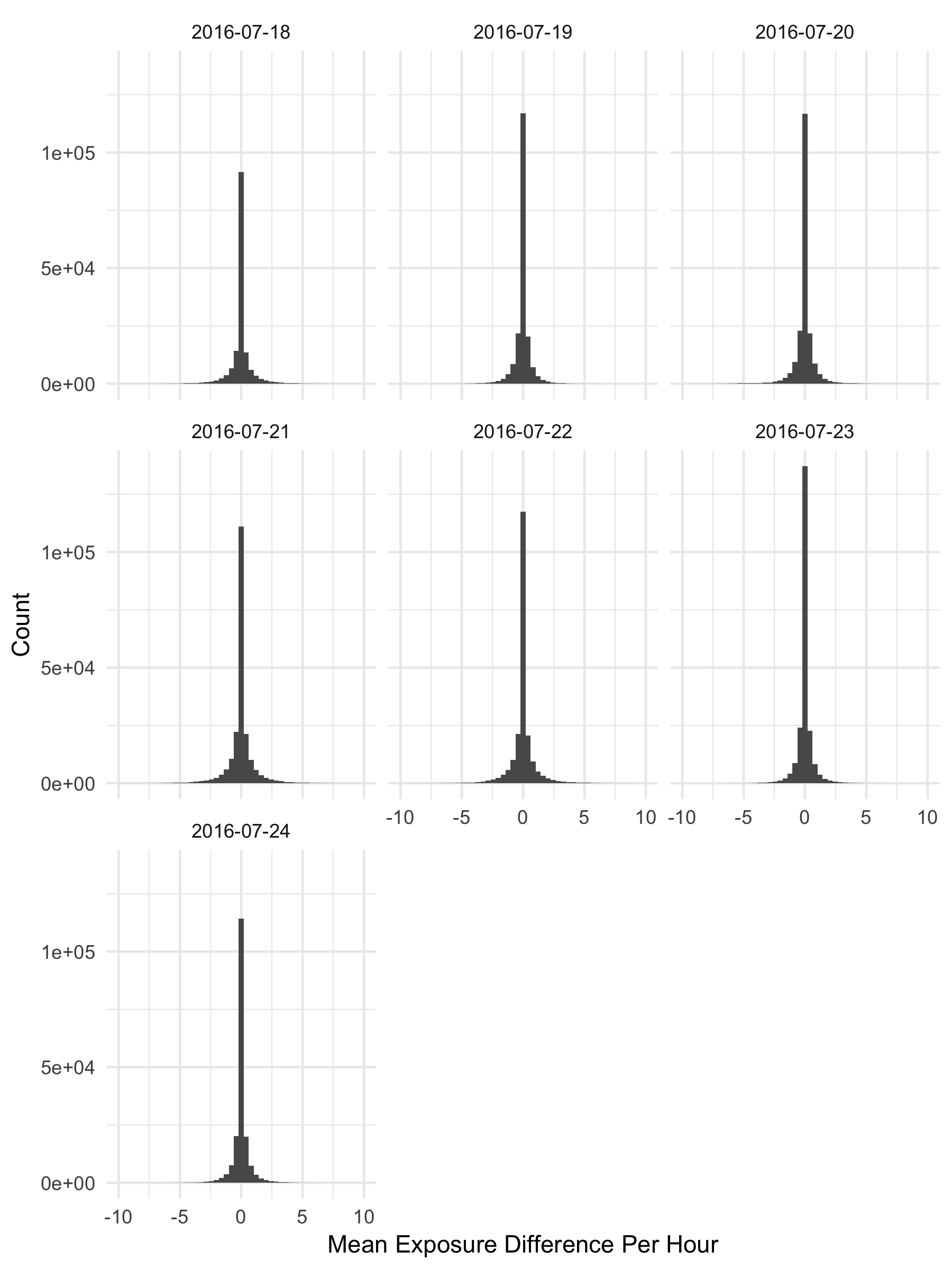}
\caption{Distribution of average hourly exposure difference (dynamic - static) based on 24-hr cumulative exposure.}
\label{fig:hourly_exp_diff_cum}
\end{figure}

Figure \ref{fig:hourly_exp_diff_cum} shows histograms (by day) of the differences in the hourly average exposure to \ozone assigned to individuals between the dynamic and static scenarios based on 24-hr cumulative exposure.  The visualization shows that the difference for each day is centered around zero and most of the absolute differences are less than 5 ppb/hour indicating that current models, not taking mobility into account, are fairly accurate. 

\begin{figure}[htbp!]
\centering
\includegraphics[width=\textwidth]{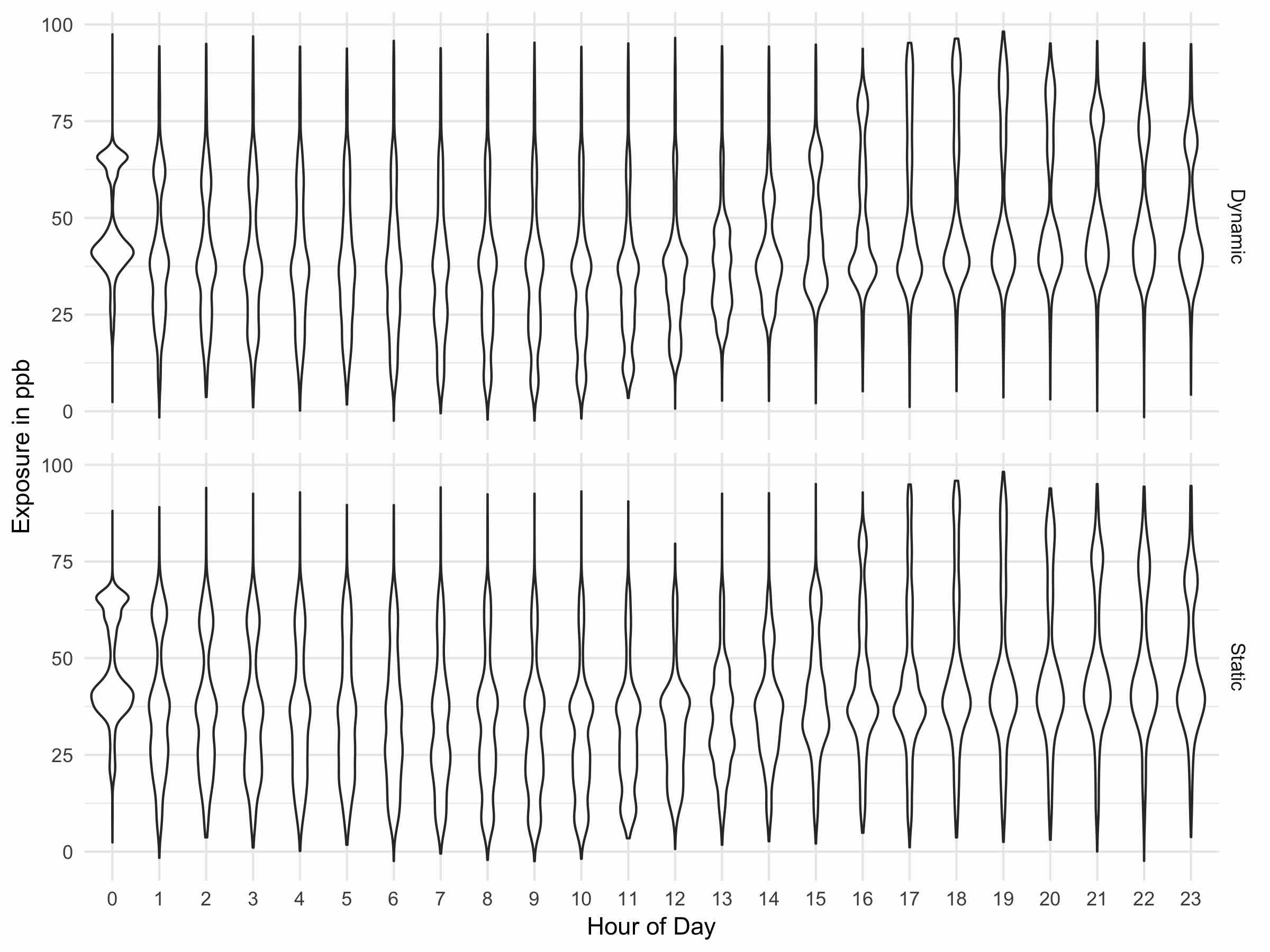}
\caption{Distribution of hourly exposure assignment using the dynamic and static scenarios.}
\label{fig:hourly_exp_dynamic_static}
\end{figure}

Figure \ref{fig:hourly_exp_dynamic_static} shows the distribution of hourly \ozone exposure assignment using the dynamic and static scenarios. The distributions look fairly similar. The static scenario generally has greater mass at lower exposures as compared to the dynamic scenario.

\pagebreak

\subsection*{Cumulative Exposure Density by Hour}

\begin{figure}[htbp!]
\centering
\includegraphics[width=\textwidth]{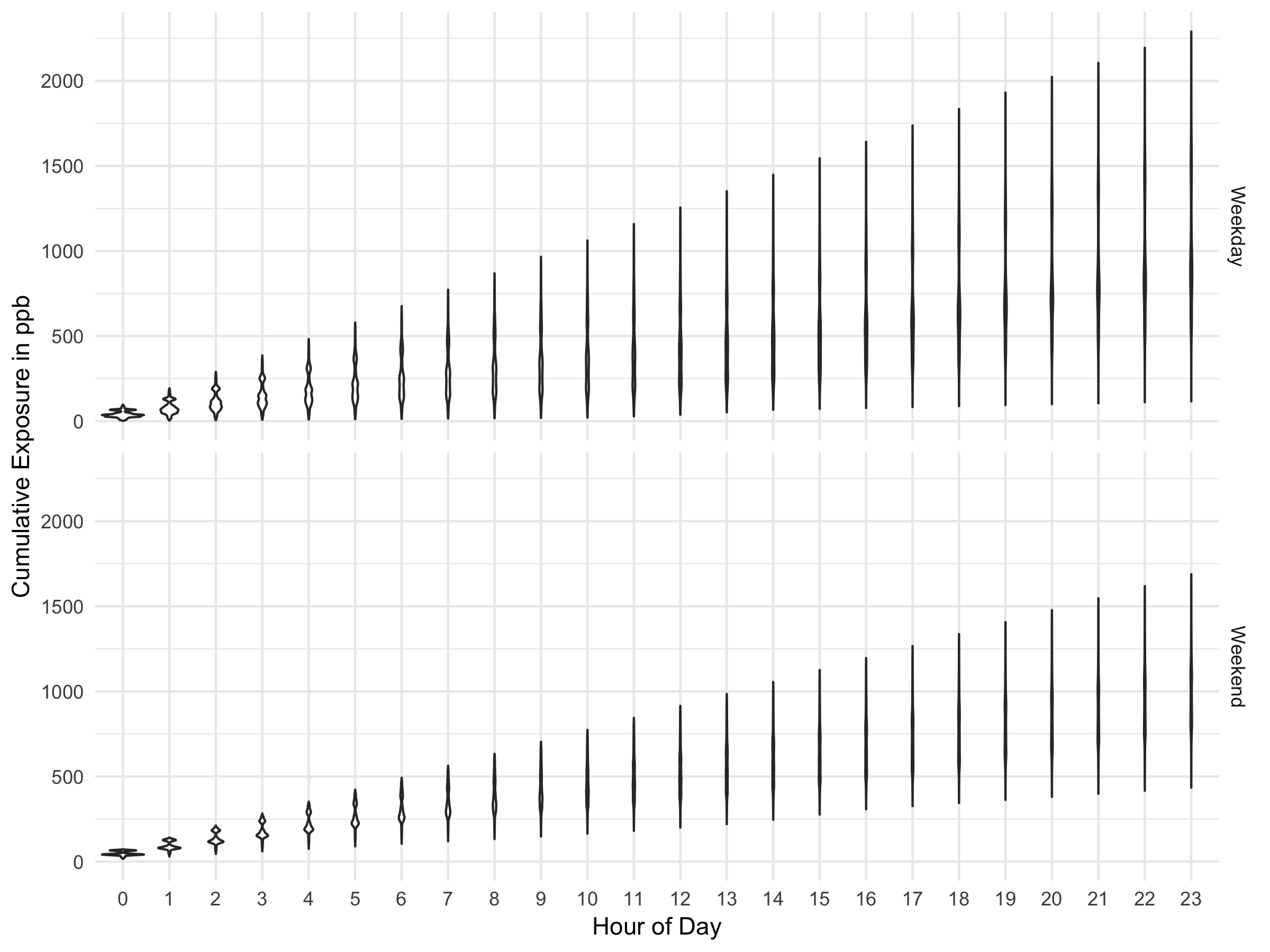}
\caption{Violin plots of the cumulative exposure distribution by hour of day for weekdays and weekends.}
\label{cum_exposure_density}
\end{figure}

Figure \ref{cum_exposure_density} shows the distribution of the cumulative \ozone exposure by hour of day for weekdays and weekends. It is interesting to note that on weekdays, the lower tail remains close to 0 ppb as the day progresses, while on weekends, the lower tail quickly moves away from 0 ppb. The distribution for each hour is more spread out on weekdays as compared to weekends.  It can also be seen that many of the individual hour exposures are bimodal, especially during the hours of 1 PM and 10 PM. Since the diffusion of ozone is relatively dispersed,, individuals in higher exposure areas likely stay in those areas (and vice versa), which would result in a large amount of variation of cumulative exposure by day.

\pagebreak

%In both plots, there appears to be a bimodal distribution for cumulative \ozone exposure.

\subsection*{Cumulative Exposure Density}

\begin{figure}[htbp!]
\centering
\includegraphics[width=\textwidth]{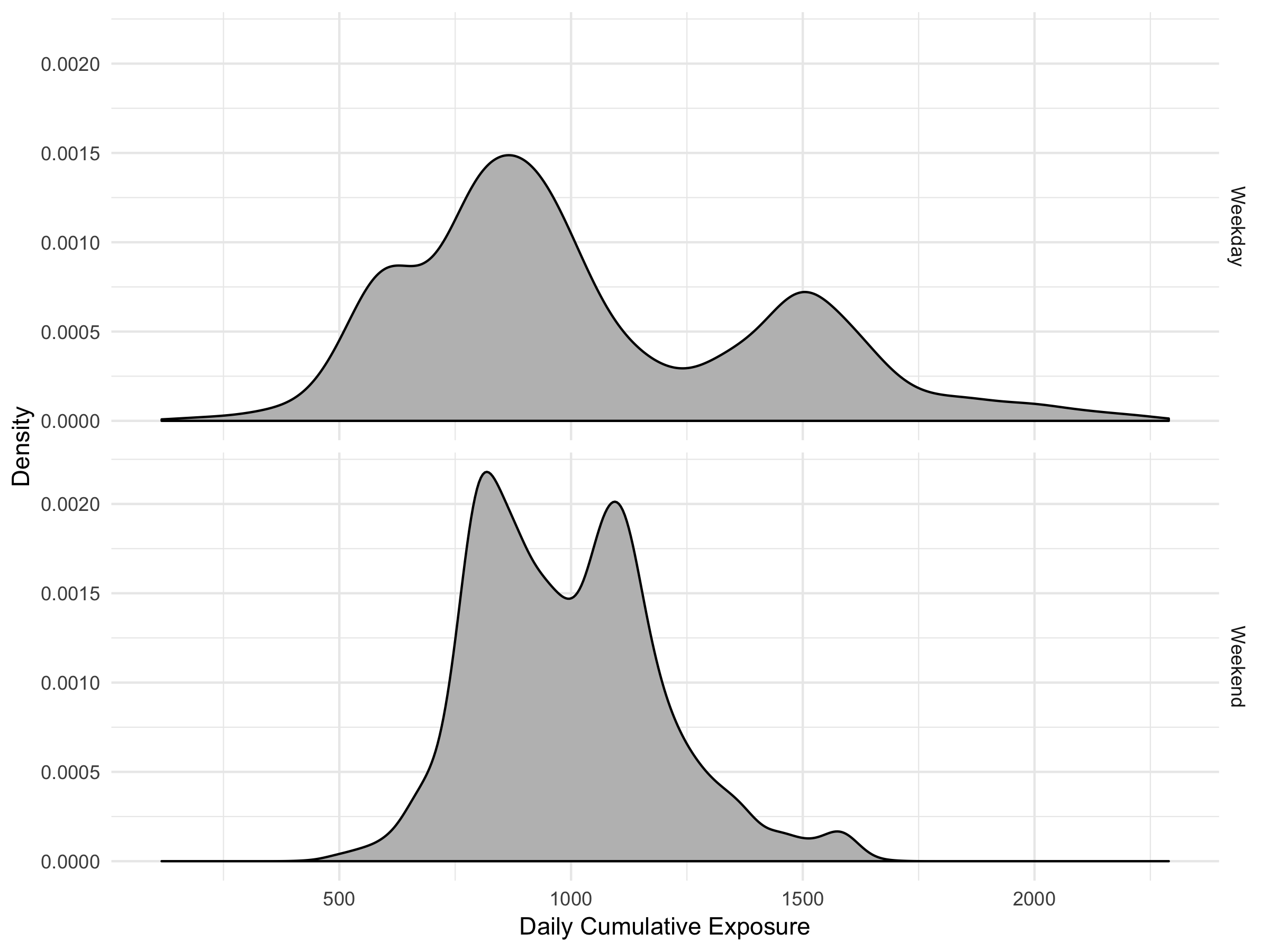}
\caption{Density plots of the daily cumulative exposure distribution.}
\label{cum_exposure_density_day}
\end{figure}

Figure \ref{cum_exposure_density_day} shows the distribution of the cumulative \ozone exposure at the end of the day for weekdays and weekends (the last hour from Figure \ref{cum_exposure_density}). The weekday distribution is more dispersed when compared to that of the weekend. This likely is a result of increased mobility, due to commuting.

%To better capture the variations in the daily, cumulative exposure, Figure \ref{cum_exposure_density_day} shows the individual-level cumulative exposure for weekdays and weekends. 

% Figure \ref{cum_exposure_density} illustrates this with violin plots of the individual-level cumulative exposure densities.

%% Authors are advised to submit their bibtex database files. They are
%% requested to list a bibtex style file in the manuscript if they do
%% not want to use model1-num-names.bst.

%% References without bibTeX database:

% \begin{thebibliography}{00}

%% \bibitem must have the following form:
%%   \bibitem{key}...
%%

% \bibitem{}

% \end{thebibliography}

\end{document}